\documentclass[12pt]{iopart}
\usepackage{iopams}
\usepackage{graphicx}
\begin{document}
\title[Exact RG equation for lattice models]{Exact renormalization group
equation for lattice Ginzburg-Landau models adapted to the solution in the 
local potential approximation}

\author{V. I. Tokar$^{1,2}$}

\address{$^1$Universit\'e de Strasbourg, CNRS, IPCMS, UMR 7504,
F-67000 Strasbourg, France}
\address{$^2$G. V. Kurdyumov Institute for Metal Physics of the 
N.A.S.\ of Ukraine, 36 Acad. Vernadsky Boulevard, UA-03142 Kyiv, Ukraine}
\begin{abstract}
The Wilson Green's function approach and, alternatively, Feynman's
diffusion equation and the Hori representation have been used to derive
an exact functional RG equation (EFRGE) that in the course of the RG flow
interpolates between the interaction part of the lattice Ginzburg-Landau
Hamiltonian and the logarithm of the generating functional of the
S-matrix. Because the S-matrix vertices are the amputated correlation
functions of the fluctuating field, it has been suggested that in the
critical region the amputation of the long-range tails makes the S-matrix
functional more localized and thus more amenable to the local potential
approximation (LPA) than the renormalized free energy functional used
in Wilson's EFRGE.

By means of a functional Legendre transform the S-matrix EFRGE has been
converted into an EFRGE for the effective action (EA). It has been found
that the field-dependent part of EA predicted by the equation is the
same as calculated within the known EA EFRGE approaches but in addition
it is accurately accounts for the field-independent terms. These are
indispensable in calculation of such important quantities as the 
the specific heat, the  latent heat, etc.

With the use of the derived EFRGE a closed expression for the
renormalization counterterm has been obtained which when subtracted
from the divergent solution of the Wetterich equation would lead to a
finite exact expression for the EA thus making two approaches formally
equivalent.

The S-matrix equation has been found to be simply connected with a
generalized functional Burgers' equation which establishes a direct
correspondence between the first order phase transitions and the shock
wave solutions of the RG equation.

The transparent semi-group structure of the S-matrix RG equation makes
possible the use of different RG techniques at different stages of the
RG flow in order to improve the LPA solution.
\end{abstract} 
\noindent{\it Keywords\/}: exact renormalization group equations,
local potential approximation, $n$-vector spin-lattice models, Burgers'
equation, shock wave solutions, first order phase transitions

\maketitle
\section{Introduction}
Exact functional renormalization group equations (EFRGEs) were
introduced by Wilson \cite{wilson} as a prospective method of dealing
with problems that cannot be solved by other techniques. An important
problem of this kind is the solution of models with strong coupling
between the fluctuating fields for which rigorous solution methods
similar to the perturbative techniques of the weak coupling case
are unavailable.  The EFRGE derived in \cite{wilson}, however,
was too complicated to be solved beyond the perturbation theory.
Therefore, using the flexibility of the RG approach, simpler EFRGEs
were derived in \cite{wegner_renormalization_1973,nicoll_exact_1976,%
polchinski_renormalization_1984,1984,maxwell_construction,%
wetterich_exact_1993,BONINI1993441,morris1994} and used in the solution
of various field-theoretic problems (see review papers
\cite{BAGNULS200191,berges_non-perturbative_2002,RG2002review,%
dupuis_nonperturbative_2021} and references to earlier literature
therein).

To the purposes of the present study the most important
will be two kinds of EFRGEs : those dealing with the
renormalization of the interaction functional suggested in
\cite{1984,polchinski_renormalization_1984} and the equations introduced
in \cite{wetterich_exact_1993,BONINI1993441,ellwanger1993,morris1994}
which renormalize the effective action (EA). Our interest to these
two approaches stems from the fact that they proved to be successful
in application to the lattice models of the Ginzburg-Landau type
\cite{latticeRG2010,CAILLOL2012291,sc-lpa-rg,tokar2019effective,tokar21BC}.
Moreover, when applied to the same problem and
solved in the local potential approximation (LPA)
\cite{local_potential,BAGNULS200191,berges_non-perturbative_2002,%
Bervillier2008525,dupuis_nonperturbative_2021,caillol_non-perturbative_2012,%
latticeRG2010,CAILLOL2012291,maxwell_construction,1984,sc-lpa-rg}
EFRGEs of both kinds produce very similar results provided that the
same cutoff in the momentum space in the form of the step function
\cite{1984,latticeRG2010} is being used in implementing the LPA. This
suggests that both kinds of equations may be equivalent in some sense
and so will give similar results in any approximation.

Another possible explanation is that LPA is very accurate and all EFRGEs
would give similar solutions if solved within this approximation.
This, however, does not seem to be plausible because when
LPA is used with RG equations that do not belong to the above two
categories, the results may disagree significantly as, e.g., between
the solutions of the Blume-Capel model in \cite{parola_recent_2012}
and in \cite{tokar21BC}. Besides, a rigorous assessment of the LPA
accuracy meets with serious difficulties. The derivative expansion
(DE) that is usually invoked as a way of justifying and correcting
LPA (see extensive bibliography on the subject in \cite{Morris_1999}
and \cite{dupuis_nonperturbative_2021}) is not a systematic approach
\cite{Morris_1999}.  For the purposes of the present study the most 
important conclusion of \cite{Morris_1999} was that DE can be valid
only at moderate interaction strengths. This was confirmed in
\cite{CAILLOL2012291} by comparison of LPA solution using Wilson-type
cutoff with the Monte Carlo (MC) simulations of \cite{hasenbusch_phi4}
where the agreement did worsen with the growing interaction strength.
But the spin-lattice models we will be interested in formally correspond
to the case of infinitely strong coupling and though in the course
of the RG flow it considerably weakens, in the transient region
\cite{wilson} the large values may persist long enough to make the
resort to DE ungrounded.  In view of this, the remarkable accuracy
of the values of the phase transition temperatures obtained within
LPA in \cite{maxwell_construction,latticeRG2010,CAILLOL2012291,%
parola_recent_2012,sc-lpa-rg,tokar2019effective,tokar21BC} in strongly
coupled lattice models requires explanation.

The aim of the present paper is to clarify the above issues and to
develop techniques facilitating the use of nonperturbative RG methods
in the solution of strongly coupled lattice models. Of special interest
to us will be the spin-lattice case because of the prominent role that
the Ising model (IM), the classical Heisenberg model, the Blume-Capel
model and many others play in statistical physics.

Specifically, in this paper we will derive EFRGE for the renormalization
of the interaction functional \cite{1984,polchinski_renormalization_1984}
using a general cutoff or, equivalently, the regulator function
\cite{dupuis_nonperturbative_2021} and will show that a fully renormalized
interaction functional coincides with the generating functional of the
logarithm of the S-matrix with minus sign. Therefore, RG equations
of this kind will be called the S-matrix equations to distinguish
them from the equations of \cite{wetterich_exact_1993,morris1994,%
berges_non-perturbative_2002,dupuis_nonperturbative_2021} which will be
called the EA equations.  We will show that EFRGEs of the two kinds are
connected by a functional Legendre transform (LT) and thus formally are
equivalent, but with a caveat.

In Wilsonian RG approach it is habitual to neglect the
field-independent multiplicative factors that contribute only to the
normalization of the field probability distribution \cite{wilson}
or, equivalently, to the field independent terms (f.i.t.) in
the renormalized Hamiltonian. According to \cite{morris1994},
this practice was also accepted in the derivation of EA EFRGEs in
\cite{wetterich_exact_1993,morris1994,berges_non-perturbative_2002,%
dupuis_nonperturbative_2021} which means that only the field-dependent
part of the EA can be assuredly correct when calculated with the use
of the EA EFRGE schemes of the above references. In many cases this is
completely sufficient which explains the successful use of the RG equation
of \cite{wetterich_exact_1993,morris1994,berges_non-perturbative_2002,%
dupuis_nonperturbative_2021} in quantitative description of the
equation of state of magnetic systems \cite{PhysRevLett.77.873} and to
the calculation of the magnetization and the critical temperatures in
lattice systems \cite{latticeRG2010,CAILLOL2012291}. The reason was that
only the field derivatives of EA were needed in the calculations and so
the f.i.t.\ disappeared from all expressions.

However, in other physically important cases the knowledge of the f.i.t.\
is indispensable. For example, the Onsager solution of the 2D IM is a
f.i.t.; the specific heat and the latent heat of the first order phase
transitions (FOPT) (see, e.g., \cite{tokar21BC}) both depend on f.i.t..

An advantage of the EA EFRGE derivation by the LT of the
S-matrix equation is that no normalization factors have been
omitted which has made possible to correct the RG approaches of
\cite{wetterich_exact_1993,morris1994,berges_non-perturbative_2002,%
dupuis_nonperturbative_2021} in such a way that they treated
the f.i.t.\ exactly.  This may have important consequences
for the theories based on the Wetterich equation in which
f.i.t.\ at the lower limit of integration is divergent
\cite{wetterich_exact_1993,morris1994,berges_non-perturbative_2002}.
To regularize the divergence various ways of
choosing the ultraviolet cutoff have been suggested in
\cite{Berges1997CoarseGA,latticeRG2011-1,latticeRG2011-2,latticeRG2012,%
cutoff2013} which, however, relied on heuristic {\em ad hoc} assumptions
which are difficult to rigorously justify.  To overcome this difficulty,
by analogy with the renormalization in the relativistic quantum field
theories a divergent counterterm will be inserted into the initial
condition of the Wetterich equation which will render f.i.t.\ in the EA
to be finite and unambiguously defined.

Besides the latent heat, discontinuities in the order parameter are
important characteristics of FOPT. The analytic structure of the S-matrix
EFRGE can be helpful in their description because when differentiated
w.r.t.\ the field variable the equation may be qualitatively described
as the generalized functional Burgers' equation (BE) \cite{gbe}.  In the
LPA it can be used to describe FOPT as the shock wave solutions of the
$n$-vector Ginzburg-Landau model for all natural $n$ in contrast to the
EA equation where the shock waves were found only in the $n\to\infty$
limit \cite{grossi2019resolving}.  The shock-wave picture puts on a
firmer footing the mechanism of the suppression of the van der Waals
loops plaguing the mean field (MF) theories discovered in the LPA RG
approach in \cite{maxwell_construction} .

In the present study of major interest will be the RG description of
non-universal quantities, such as the phase transition temperatures,
because of their truly non-perturbative nature. Universal quantities,
such as the critical exponents, though not accurately predicted by the
LPA do not depend on the interaction strength and in the three-dimensional
systems that we will consider in this paper can be efficiently calculated
by perturbative techniques within simplified Hamiltonians characterized
by a few parameters \cite{wilson,Pelissetto2002CriticalPA}. In contrast,
non-universal quantities may depend on an unlimited number of parameters
and are particularly difficult to calculate beyond the perturbation
theory.

Obviously, the universal quantities are also needed for a comprehensive
description of critical phenomena in strongly interacting systems so
it would be reasonable to accommodate the available rigorous results
\cite{wilson,Pelissetto2002CriticalPA} to the LPA solutions.  To this
end in the present paper a multi-step renormalization technique will be
developed that will make possible the use of different renormalization
methods at different stages of the RG flow. For example, it will be
possible to use LPA within the transient region at an early stage of
renormalization when the interactions are strong \cite{wilson} and to
switch to the perturbative treatment in the critical region where the
conventional perturbation theory or the DE expansion become efficient
\cite{Morris_1999,PhysRevE.101.042113,shalaby_critical_2021}.

The presentation of material will be as follows. After introducing
the necessary notation in the next section, in section \ref{efrges}
we will derive the EFRGE for the interaction functional that underlies
the LPA equations of \cite{1984,tokar2019effective,sc-lpa-rg,tokar21BC}.
It will be shown that the RG flow interpolates between the initial local
potential and the generating functional of the logarithm of the S-matrix.
It will be argued that in conjunction with a self-consistency (SC)
condition the S-matrix functional is more local than the FE functional of
the Wilson approach \cite{wilson} so the S-matrix EFRGE should be more
accurately represented in the LPA than the Wilson one \cite{wilson}. In
\ref{legendre} several forms of a LT will be introduced that connect
the S-matrix and the EA RG equations both the exact and in the LPA form
including the case of the multi-step renormalization.  Because EFRGEs
are the evolution equations, their solutions depend on the initial
conditions.  In sections \ref{phi4}, \ref{spin-lattice} and \ref{ini} it
will be shown that for S-matrix EFRGE analytic initial conditions can be
easily established even in the spin-lattice case despite that it formally
corresponds to an infinitely strong coupling. The renormalization of the
divergence in the Wetterich equation will be discussed in \ref{wetterich}.

In section \ref{feynman-eq} we will show that the Hori representation
makes the (semi)group structure of the S-matrix EFRGE fully transparent
which simplifies the stepwise renormalization needed in the preliminary
exact renormalization in the spin models and also allows one to switch
to a perturbative technique in the critical region.

Further, in section \ref{lpa} qualitative arguments will
be given in support of the standpoint that the use of
the step function cutoff for the elimination of high momenta
\cite{1984,latticeRG2010,CAILLOL2012291,tokar2019effective,sc-lpa-rg,tokar21BC}
should be more accurate in the lattice case than in the continuum models.
The FOPT will be discussed in section \ref{fopts} and in \ref{irim}.

In final section \ref{discuss} a brief summary will be given. 
\section{\label{defs}Definitions and notation} 
To simplify notation and to facilitate comparison with \cite{wilson} all
derivations in the present paper will be done for the Landau-Ginzburg
model of the Ising universality class, that is, with the scalar field
variable. Generalization to the $n$-vector models can be achieved along
the same lines as in \cite{1984,tokar2019effective} and in simple cases
is straightforward (see section \ref{spin-lattice} below).

Thus, we will be interested in the calculation of the partition functional
\begin{equation}
Z(\vec{h})=\int d\vec{s}\, \rme^{-H_0(\vec{s})+\vec{h}^\dagger\vec{s}}
\label{Z}
\end{equation}
by means of the $N$-dimensional integral over the fluctuating scalar
field which will be usually denoted as a column vector $\vec{s}=[s_i]$
(similarly, $\vec{h}=[h_i]$) defined at sites $i$ of a periodic lattice
of size $N$; $\int d\vec{s}\equiv\prod_i\int_{-\infty}^\infty ds_i$. The
factor $1/k_BT$ will be assumed to be included in the parameters of
the dimensionless Hamiltonian of the Landau-Ginzburg model 
\begin{equation}
\fl H_0(\vec{s}) = \frac{1}{2}\sum_{ij}\left(\epsilon_{ij}
+r\delta_{ij}\right)s_is_j +U_0(\vec{s})
=\frac{1}{2}\vec{s}\,^\dagger(\hat{\epsilon}+r\hat{I})\vec{s}+U_0(\vec{s})
\label{H0}
\end{equation}
where $\hat{\epsilon}=[\epsilon_{ij}]$ is the matrix of the pair
interactions which in the thermodynamic limit $N\to\infty$ (which will be
always implicitly assumed) is translationally invariant; the strength of
the pair interactions will be characterized by a dimensionless constant $K$
with $\hat{\epsilon} \propto K$; $\hat{I}$ is the unit matrix, $U_0$
the interaction potential and $\vec{h}$ is the source and/or the
external field.

The dagger symbol used in (\ref{H0}) denotes Hermitian conjugation which
for real fields means the transposition but for the Fourier transformed
fields the complex conjugation should be additionally applied.

For convenience we will use The symmetric Fourier transform, e.g., in
\begin{equation}
s_{\bf k}=N^{-1/2}\sum_j\rme^{-\rmi{\bf k}\cdot j}s_j
= \sum_j\left(\hat{{\cal F}}\right)_{{\bf k},j}s_j; 
\label{fourier}
\end{equation}
the use of the unitary matrix $\hat{{\cal F}}$ ($\hat{{\cal
F}}^{-1}=\hat{{\cal F}}^\dagger$) will allow us to formally consider
the real-space vector and its Fourier transform as the same abstract
vector in two reference frames connected by a unitary rotation. This
makes possible the use of the same symbol for vectors both with the
lattice and with the momentum components.

The bare interaction $U_0$ in (\ref{H0}) can in principle be any
functional of $\vec{s}$, but in the present paper we will assume the
interaction functional to be the sum of site-local potentials 
\begin{equation}
U_0(\vec{s})=\sum_iu(s_i,0)
\label{U0}
\end{equation}
where the second argument of $u$ is $t$,---a scalar variable that will
parametrise the RG flow. It can be chosen arbitrarily so we define it to
be varying from $t=0$ corresponding to the system in its initial or
``bare'' state to $t=t^{\mathrm R}$ in the fully renormalized state. The
bare potential in (\ref{U0}) will be assumed to be analytic in the field
variable but as we will see later the analyticity of the renormalized
potential at $t=t^{\mathrm R}$ can be broken by phase transitions.

Here and below by superscript ``R'' we will mark all fully renormalized
quantities that appear also in only partially renormalized form at arbitrary
$t$.

Further, because the separation of Hamiltonian in the quadratic and the
interaction parts is not unique, the ambiguity has been used to define
the first term in (\ref{H0}) in such a way that, first, the Fourier
transform of $\hat{\epsilon}$ had the small-momentum asymptotic
\begin{equation}
\epsilon({\bf k})_{k\to0}\simeq ck^2
	\label{k20}
\end{equation}
where $k=\vert{\bf k}\vert$ and $c$ is a constant proportional to $K$.
This can be achieved by subtracting an appropriate site-diagonal term
in the quadratic part of (\ref{H0}) and adding it to $U_0$. Second,
an arbitrary diagonal term $r\hat{I}$ has been added to the quadratic
part and subtracted from $U_0$.  Obviously, neither $H_0$ nor the exact
partition function (\ref{Z}) depend on $r$ but in an approximate solution
the independence may be broken and the arbitrary parameter can be used
to improve the accuracy.  This will be done below with the use of a SC
condition (section \ref{sc}).

The quantities that we are going to calculate in this paper will be the 
Helmholtz FE
\begin{equation}
F^{\mathrm R}(\vec{h})=-\ln Z(\vec{h})
\label{FE}
\end{equation}
and its two derivatives: the magnetisation
\begin{equation}
	{m}_{ i} \equiv \langle s_{ i}\rangle
	=-\frac{\partial F^{\mathrm R}({\vec{h}})}{\partial {h_{ i}}},
	\label{s_av}
\end{equation}
and the pair correlation function
\begin{equation}
	G_{ij}^{\mathrm R}= -\frac{\partial F^{\mathrm R}({\vec{h}})}{\partial {h_{ i}}
\partial {h_{ j}}}
=\langle {s}_{ i}{s}_{ j}\rangle -
{m}_{ i}{m}_{ j}.
	\label{ss_av}
\end{equation}

Throughout the paper the arrows will denote the $N$-dimensional lattice
vectors; Fourier momenta and $n$-vectors will be boldface and for
simplicity the same letters but without arrows or subscripts, such as
$h$ and $m$, will denote the scalar values of the homogeneous external
field and of the on-site magnetisation, respectively. In particular,
according to (\ref{fourier}) the homogeneous external field will have
vector components
\begin{equation}
h_{\bf k}^h=\sqrt{N}\delta_{\bf k,0}h.
\label{h}
\end{equation}
\section{\label{efrges}Exact RG equations}
In the derivation of S-matrix EFRGE we will use the Wilson
Green's function approach (see \cite{wilson}, ch.\ 11)
which will be slightly modified in order to facilitate
comparison with the EA approach as presented in review articles
\cite{berges_non-perturbative_2002,dupuis_nonperturbative_2021}.
To begin with let us consider a$N$-dimensional Gaussian kernel
\begin{equation}
{\cal G}(\vec{s},\vec{s}\,^\prime,t)={\det{[\hat{{R}}(t)}/{2\pi}]}^{\frac{1}{2}}
\exp \left[-\frac{1}{2}\left( \vec{s}-\vec{s}\,^\prime \right)^{\scriptscriptstyle T}
\hat{{R}}(t)\left( \vec{s}-\vec{s}\,^\prime \right)\right]
\label{green}
\end{equation}
where $\hat{R}(t)$ is the regulator matrix
\cite{berges_non-perturbative_2002,dupuis_nonperturbative_2021} which
is assumed to be a non-negative definite, symmetric, translationally
invariant and that its site-diagonal matrix elements behave at $t\to0$ as
\begin{equation}
\hat{{R}}(t\to0)\to\infty
\label{R2infty}
\end{equation}
in order to satisfy the conventional initial condition
\begin{equation}
{\cal G}(\vec{s},\vec{s}\,^\prime,t=0)=\delta(\vec{s}-\vec{s}\,^\prime)\equiv\prod_i\delta(s_i-s_i^\prime).
\label{Gt=0}
\end{equation}

In the Wilson approach the regulator $\hat{R}$ in (\ref{green}) should be
chosen in such a way that the convolution of the Green function with the
Boltzmann factor $\rme^{-H_0+\vec{h}^\dagger\vec{s}}$ at $t^{\mathrm R}$
satisfied 
\begin{equation}
Z(\vec{h})=C\int d\vec{s}\,^\prime{\cal G}
(\vec{s},\vec{s}\,^\prime,t^{\mathrm R})\rme^{-H_0(\vec{s}\,^\prime)
+\vec{h}^\dagger\vec{s}\,^\prime}
\label{convolution}
\end{equation}
with some proportionality coefficient $C$ which may depend on $\vec{s}$,
$\vec{h}$, $t^{\mathrm R}$, etc., with the only requirement that it
could be calculated exactly.

Replacing $t^{\mathrm R}$ in (\ref{convolution}) by $t\in (0,t^{\mathrm
R})$ one obtains an expression that interpolates between the Boltzmann
factor and the fully renormalized partition function. Because the Green
function satisfies a diffusion-type differential equation, by choosing
a suitable form of the regulator one can derive a Wilsonian EFRGE in
differential form, as shown in \cite{wilson} with a particular choice
of the regulator. In this paper we will derive along similar lines
a different, S-matrix EFRGE, adopted for the use in conjunction with
the LPA.

But before proceeding it is pertinent to point out that similar to
the Wilson equation \cite{wilson}, the S-matrix EFRGE deals with the
calculation of the Helmholtz FE. On the other hand, the EA in application
to statistical-mechanical problems aims at calculating the Gibbs FE
which necessitated the use of a modified LT in the
derivation of EA EFRGE \cite{dupuis_nonperturbative_2021}. Of course,
the two fully renormalized thermodynamic potentials are connected by
the conventional LT \cite{zia}. 
\subsection{S-matrix EFRGE}
The advantage of S-matrix EFRGE derived below
in comparison with the Wilson-type RG equations
\cite{wilson,wegner_renormalization_1973,nicoll_exact_1976}
and with EA EFRGEs of
\cite{wetterich_exact_1993,BONINI1993441,ellwanger1993,morris1994} is
that from a qualitative standpoint they are not suitable for the LPA in
lattice models even with local interaction potentials.  In both cases
EFRGEs deal with renormalization of the total Hamiltonian which contains
both the local potential and the non-local quadratic part. Because EFRGEs
are essentially nonlinear, in the course of the RG flow the two terms
are intermixed which, in particular, spreads the interaction potential
over many lattice sites thus making the LPA along the whole RG trajectory
dubious from a qualitative point of view.

In such cases LPA can be justified only formally within the DE
\cite{Morris_1999,dupuis_nonperturbative_2021}. However, for DE
to converge the effective expansion parameter should be at least
smaller than unity. This is possible in the continuum models with
a single cutoff parameter $\Lambda$. The lowest order correction in
the dimensionless DE parameter in the isotropic systems will be of
order $(k/\Lambda)^2\le1$. In the lattice case, however, the cutoff is
determined by the lattice constant $a_{\mathrm lat}$ but because in the
corners of the Brillouin zone (BZ) of, e.g., sc lattice the squared
momentum can be as large as $(ka_{\mathrm lat})^2\approx 3\pi^2\gg1$
the use of DE becomes meaningless at the early stages of the RG flow.
(Henceforth we chose $a_{\mathrm lat}$ to be our length unit.)

Therefore, to justify LPA in lattice models at least qualitatively it
seems reasonable to derive an EFRGE for the interaction part of the
Hamiltonian and develop approximations that would keep the renormalized
potential during the RG flow  as local as possible. To this end let us
consider functional
\begin{equation}
S(\vec{s},t)=\rme^{-U(\vec{s},t)}
\label{S-U}
\end{equation}
subject to the initial condition $U(\vec{s},t=0)=U_0(\vec{s})$.  In the Wilson
Green's function approach the RG evolution of $S$ is described by the equation
\begin{equation}
S(\vec{s},t)=\int {\cal G}(\vec{s},\vec{s}\,^\prime,t)
\rme^{-U_0(\vec{s}\,^\prime)}\,d\vec{s}\,^\prime.
\label{S(t)}
\end{equation}
From explicit form of ${\cal G}$ (\ref{green}) it is easy to surmise
that the partition function can be obtained from the fully renormalized
$S^{\mathrm R}$ or, equivalently, from $U^{\mathrm R}$ if instead of
(\ref{R20}) we impose on the regulator the following condition
\begin{equation}
\hat{{R}}^{\mathrm R}=\hat{R}(t^{\mathrm R})=\hat{\epsilon}+r\hat{I}.
\label{RtR}
\end{equation}
With this regulator one gets at $t=t^{\mathrm R}$ 
after expanding the quadratic form in (\ref{green}) 
\begin{equation}
\rme^{-U^{\mathrm R}(\vec{s})}=
\sqrt{\det{\left(\hat{{R}}^{\mathrm R}/{2\pi}\right)}}
\exp\left(-\frac{1}{2}\vec{s}\hat{{R}}^{\mathrm R}\vec{s}\right)Z(\hat{{R}}^{\mathrm R}\vec{s}).
\label{psi-tR}
\end{equation}
Next choosing $\vec{s}$ to satisfy
\begin{equation}
\hat{{R}}^{\mathrm R}\vec{s}=\vec{h}
\label{Rs=h}
\end{equation}
we obtain the partition functional 
\begin{equation}
Z(\vec{h}) = \sqrt{\det{\left({2\pi}\hat{{G}}^{\mathrm P}\right)}}
\exp\left(\frac{1}{2}\vec{h}\hat{{G}}^{\mathrm P}\vec{h}\right)
\rme^{-U^{\mathrm R}(\hat{{G}}^{\mathrm P}\vec{h})}
\label{ZR}
\end{equation}
where 
\begin{equation}
\hat{{G}}^{\mathrm P}=\left(\hat{{R}}^{\mathrm R}\right)^{-1}
=\left(\hat{\epsilon}+r\hat{I}\right)^{-1}
\label{GP}
\end{equation}
is the bare propagator. The superscript ``P'' has been used instead of
``R'' to distinguish $\hat{{G}}^{\mathrm P}$ from the fully renormalized
pair correlation function $G^{\mathrm R}_{ij}$ defined in (\ref{ss_av}).

The FE functional (\ref{FE}) corresponding to (\ref{ZR}) reads
\begin{equation}
F^{\mathrm R}(\vec{h}) = U^{\mathrm R}(\hat{G}^{\mathrm P}\vec{h})
-\frac{1}{2}\vec{h}^\dagger\hat{{G}}^{\mathrm P}\vec{h}
-\frac{1}{2}\Tr\ln{\left(2\pi\hat{{G}}^{\mathrm P}\right)}.
\label{FH}
\end{equation}
\subsection{\label{sc}Self-consistency condition} 
Standard perturbative analysis (see, e.g., \cite{wilson,vasiliev1998})
shows that renormalized interaction vertices in $F^{\mathrm R}(\vec{h})$
(\ref{FH}) are the connected correlation functions coupled to the source
field $\vec{h}$ with the external legs in the diagrammatic representation
corresponding to the exact pair correlation functions $\hat{G}^{\mathrm
R}$.  Here by interaction vertices we mean the terms in $F^{\mathrm R}$
of the third and higher orders in the source field which obviously are all
contained in the renormalized potential $U^{\mathrm R}$.  But according
to \cite{wilson} (see, e.g., equation (7.45)), factors $\hat{G}^{\mathrm
R}$ describe the leading long-distance behaviour at the critical point
when the nonlocality is the most pronounced.  Thus, if instead of $\hat{
G}^{\mathrm P}$ in the argument of $U^{\mathrm R}$ in (\ref{FH}) we had
$\hat{G}^{\mathrm R}$ then functional $U^{\mathrm R}(\vec{s})$ would be more
local than $F^{\mathrm R}$ and so better representable by the LPA.

Obviously that with only one parameter $r$ at hand it is in general not
possible to make $\hat{G}^{\mathrm P}$ to be equal to $\hat{G}^{\mathrm
R}$ because in the exact pair correlation function the mass operator will
not be local in nontrivial models. However, in ferromagnetic systems
the long-distance behaviour is governed by the smallest momenta in the
Fourier transformed pair correlation function so for our purposes it
would be sufficient to satisfy a simpler requirement
\begin{equation}
	G^{\mathrm R}({\bf k\to0})\approx G^{\mathrm P}({\bf k\to0})=1/r
	\label{G=G}
\end{equation}
where according to (\ref{GP})
\begin{equation}
G^{\mathrm P}({\bf k}) =\frac{1}{\epsilon({\bf k})+r}.
\label{G(k)}
\end{equation}
It should be pointed out that here we have neglected the modification of
the long-distance behaviour due to the critical exponent $\eta$. It can
be expected, however, that it is important only when $r\to0$. Besides,
the exponent is equal to zero in the LPA so anyway we could not account
for it in this approximation. A possibility to correct this within our
RG approach will be discussed in section \ref{exponents}.

Now substituting (\ref{FH}) into (\ref{ss_av}) one finds that condition
(\ref{G=G}) in terms of the fully renormalized potential reads
\begin{equation}
\left.\frac{\partial^2U^{\mathrm R}(\vec{s})}
{\partial{s_{\bf k}}\partial{s_{\bf -k}}}
\right|_{{\bf k\to0},s_{\bf k=0}}=0 
\label{SCU}
\end{equation}
where we assume that the solution has been obtained in a
translationally-invariant system in homogeneous external field $h$
so that $\vec{s}=\hat{G}^{\mathrm P}\vec{h}$ in (\ref{FH}) in this case 
should be calculated with the use of (\ref{h}) and (\ref{G(k)}) as
\begin{equation}
s_{\bf k=0}=\left.G^{\mathrm P}({\bf k})h^h_{\bf k}\right\vert_{\bf k=0}
=\sqrt{N}h/r
\label{s0}
\end{equation}
The SC condition (\ref{SCU}) will be used everywhere below in the
solutions of the LPA RG equations.
\subsection{S-matrix EFRGE in differential form}
Differentiating (\ref{S(t)}) w.r.t.\ $t$, using in the derivative of
${\cal G}$ (\ref{green}) the commutativity of translationally-invariant
matrices and Jacobi's formula for invertible matrices one arrives at the
exact linear RG equation 
\begin{equation}
\partial_tS(\vec{s},t)=\frac{1}{2}\sum_{ij}\partial_t{{G}}_{ij}
\frac{\partial^2 S(\vec{s},t)}{\partial {s_i}\partial {s_j}}
\label{diff-eq}
\end{equation}
where $\partial_t\equiv\partial/\partial t$ and 
\begin{equation}
\hat{G}(t)=[G_{ij}(t)]=\hat{{R}}^{-1}(t).
\label{Gdef}
\end{equation}
Substituting $S$ from (\ref{S-U}) in (\ref{diff-eq})
the non-linear EFRGE for the interaction potential
in the lattice coordinates is obtained as
\begin{equation}
\partial_t{ U} = \frac{1}{2}\sum_{ij}{\partial_t{G}}_{ij}^{(0)}
\left(\frac{\partial^2 U}{\partial {s_i}\partial {s_j}}
-\frac{\partial U}{\partial {s_i}}\frac{\partial U}{\partial {s_j}}\right).
\label{wilson-eq}
\end{equation}
But in implementation of the Wilsonian renormalization we will need it
in the momentum representation \cite{1984,polchinski_renormalization_1984}
\begin{equation}
\partial_t U(\vec{s},t)+\frac{1}{2}\sum_{\bf k}\partial_t G({\bf k},t)
\frac{\partial U}{\partial s_{\bf -k}}\frac{\partial U} {\partial s_{\bf
k}} =\frac{1}{2}\sum_{\bf k}\partial_t G({\bf k},t) \frac{\partial^2
U}{\partial s_{\bf -k}\partial s_{\bf k}}. \label{exact0} 
\end{equation}
Here the terms have been rearranged in such a way that after
differentiation w.r.t.\ $s_{\bf q}$ (\ref{exact0}) acquired the structure
of a $N$-dimensional generalized BE \cite{gbe,gao_analytical_2017}
for $\vec{\cal U}_s=\left[\partial U/\partial s_{\bf q}\right]$. It
will greatly simplify under LPA and in this form will be useful in the
description of FOPTs in section \ref{fopts}.
\subsection{\label{feynman-eq}Feynman's diffusion equation as an EFRGE}
In our notation Feynman's diffusion equation \cite{hori1952} is obtained
from (\ref{diff-eq}) by assuming that $G_{ij}(t)=tG^{\mathrm P}_{ij}$,
$0\leq t\leq t^{\mathrm R}=1$:
\begin{equation}
\partial_t S(\vec{s},t)=\frac{1}{2}\sum_{ij}{G}^{\mathrm P}_{ij}
\frac{\partial^2 S(\vec{s},t)}{\partial {s_i}\partial {s_j}}.
\label{feynman}
\end{equation}
Thus, (\ref{feynman}) is an EFRGE with a particular choice of the cutoff
matrix $\partial_t\hat{G}(t)$.

The Hori representation \cite{hori1952,vasiliev1998,1984} can be obtained
by formally integrating (\ref{feynman}) or, more generally, 
(\ref{diff-eq}) as
\begin{equation}
S(\vec{s},t)=\exp\left(\frac{1}{2}\sum_{ij}{G}_{ij}(t)\frac{\partial^2 }
{\partial {s_i}\partial {s_j}}\right)S_0(\vec{s}).
\label{hori}
\end{equation}
It can also be obtained independently of Feynman's equation
\cite{vasiliev1998} so the latter as well as the S-matrix EFRGE
(\ref{exact0}) can be straightforwardly derived from (\ref{hori})
by simple differentiation w.r.t.\ $t$ \cite{1984,sc-lpa-rg}.

An important property of the Hori representation is that its semi-group
RG structure becomes completely transparent if $\hat{G}(t)$ in the
exponential is written as
\(
\int_0^t \partial_{t^\prime}\hat{G}(t^\prime)d t^\prime.
\)
Now from the properties of the integral it is seen that the
renormalization can be performed in several finite steps, for example,
first from $t=0$ to $t=t_0$ and then from $t_0$ to $t$. At the second step
$S(\vec{s},t_0)$ should be taken as the initial condition and $\hat{G}(t)$
in the exponential should be replaced by
\begin{equation}
\hat{\Delta}(t,t_0)=\hat{G}(t)-\hat{G}(t_0).
\label{Delta}
\end{equation}
This possibility will be used in sections \ref{lpa} and \ref{exponents}
below. The multi-step renormalization can be also obtained in the Green's
function approach (see \ref{steps}), though not as straightforwardly as
in the Hori representation.
\subsection{Legendre transform of S-matrix EFRGE}
In \ref{legendre} explicit LTs of the S-matrix
equation have been performed both exactly and in the LPA. It
has been shown that the EA equation (\ref{wetterich1}) from
\cite{wetterich_exact_1993,morris1994,berges_non-perturbative_2002,%
dupuis_nonperturbative_2021} can be obtained from (\ref{exact0}) by
neglecting f.i.t.\ in the exact equation. We use somewhat different
notation in these equations because it is not clear whether our
LT in \ref{legendre} and the modified LT in 
\cite{wetterich_exact_1993,morris1994,berges_non-perturbative_2002,%
dupuis_nonperturbative_2021} are the same at intermediate stages of the
RG flow. But the equations are mathematically equivalent so with the
same initial conditions will lead to the same solutions. 

The EA EFRGE (\ref{exact1}) takes into account all contributions
and so can be used in statistical-mechanical calculations.  It is to
be noted that because with condition (\ref{RtR}) $G({\bf k},t)$ in
(\ref{exact1}) is not divergent neither at $t=0$ nor at $t^{\mathrm
R}$ which makes this EFRGE easier to deal with computationally
than equations (\ref{wetterich1}) and (\ref{wetterich-eq}). It
is this EA EFRGE in LPA that was used in all calculations in
\cite{sc-lpa-rg,tokar2019effective,tokar21BC}. An equation similar to
(\ref{exact1}) was obtained in \cite{morris1994} but in the derivation
the normalization factors and the vacuum contributions that correspond to
f.i.t.\ in FE case have been dropped.
\section{\label{lpa}LPA}
The step-function cutoff suggested in \cite{1984}
for continuum models proved to be very accurate
in application to LPA solutions of lattice models
\cite{maxwell_construction,latticeRG2010,CAILLOL2012291,parola_recent_2012,%
sc-lpa-rg,tokar2019effective,tokar21BC}. This success can be
qualitatively understood within the Kadanoff picture of critical phenomena
\cite{block-spins1966,wilson} as follows.

Formally, LPA consists in assuming that functional $U(\vec{s},t)$
preserves the local structure of $U_0$ (\ref{U0}) throughout the whole
evolution from $t=0$ to $t=t^{\mathrm R}$:
\begin{eqnarray}
U(\vec{s},t)\stackrel{LPA}{\approx}N\sum_{l,\{{\bf k}_i \}} 
N^{-\frac{l}{2}}u_l(t)s_{{\bf k}_1} s_{{\bf k}_2}\dots s_{{\bf k}_l}
\delta^L_{{{\bf k}_1} +{{\bf k}_2}+\dots +{{\bf k}_l}},
\label{ULPA}
\end{eqnarray}
i.e., the Fourier-transformed coefficients of the expansion do not depend
on the momenta apart from the lattice Kronecker symbols $\delta^L$. The latter
differs from the conventional Kronecker delta in that the momenta 
may sum not only to zero but to any of the reciprocal lattice vectors, as  
can be seen from the formal definition (\ref{deltaL}). 

The LPA {\em ansatz} (\ref{ULPA}) allows one to establish a one-to-one
correspondence between functional $U$ and (in the Ising universality
class) a function of a real variable $x$
\begin{equation}
u(x,t) = \sum_{l=0}^\infty u_l(t)x^l
\label{u(xt)}
\end{equation}
which can be interpreted as the on-site potential.  The general
$n$-vector case is treated similarly with using $n$-vector ${\bf
x}=(x_1,x_2,\cdots,x_n)$ instead of $x$ \cite{1984}.

The evolution equation for $u$ should be obtained by substituting
(\ref{ULPA}) into (\ref{exact0}). However, in general case the equation
will be incompatible with the LPA {\em ansatz} so the task is to satisfy it
approximately. The difficulty causes the momentum dependence of
the cutoff function $\partial_t G({\bf k},t)$. Indeed, if it was equal
to unity
\begin{equation}
\partial_t G=1, 
\label{G_t=1}
\end{equation}
then equation (\ref{exact0}) could be satisfied exactly because in
the real space the inverse Fourier transform of unity is $\delta_{ij}$
which would make the RG equation separable, that is, representable as the
sum over the sites of identical local potentials.  Feynman's diffusion
equation (\ref{feynman}) in this case would admit solution in quadratures
which in terms of $u$ would read
\begin{equation}
	\rme^{-u({ x},t)}
=\frac{1}{\sqrt{2\pi t}}\int_{-\infty}^\infty d{ x}_0\;
\rme^{-({ x-x}_0)^2/{2t}}\rme^{-u(x_0,0)}.
\label{f_ini}
\end{equation}

\begin{figure}[htp]
\centering \includegraphics{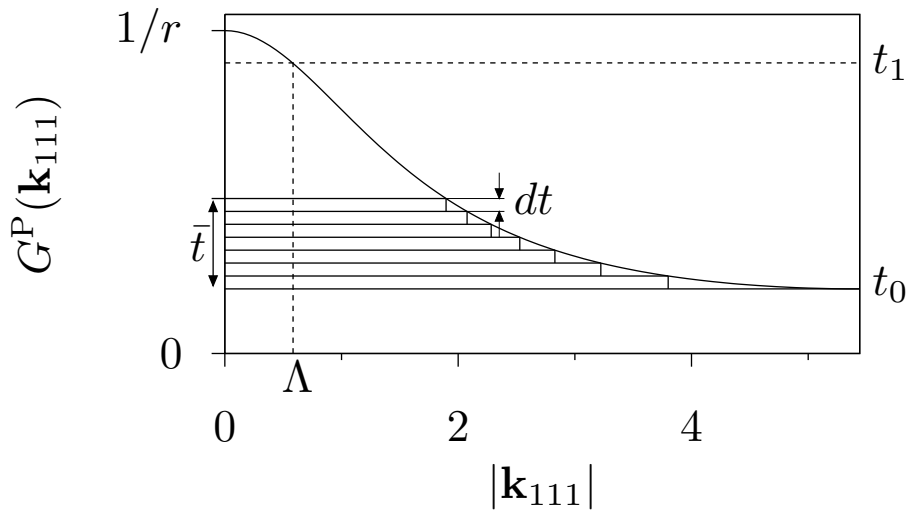}
\caption{\label{figure1} Illustration of the layer-cake renormalization
scheme. The stacked narrow rectangles are cross-sections of
$dG=\theta[G^{\mathrm P}({\bf k})-t]dt$ along the diagonal of the
Brillouin zone on which $G^{\mathrm P}({\bf k})$ reaches its minimum
value $t_0$ defined in (\ref{t0}); $\bar{t}= t-t_0$; $\Lambda$ is an
approximate cutoff momentum in the region where $G^{\mathrm P}({\bf
k})$ is almost isotropic (for further explanations see the text).}
\end{figure}

The layer-cake renormalization scheme
\cite{1984,sc-lpa-rg,tokar2019effective} has been devised with the aim
of emulating (\ref{G_t=1})--(\ref{f_ini}) with a maximum possible accuracy
in the case of momentum-dependent cutoff function.  To this end the latter
was chosen as
\begin{equation}
\partial_t G({\bf k},t) = \theta[G^{\mathrm P}({\bf k})-t]
\label{cutoff}
\end{equation}
where $\theta$ is the step function that is equal to unity when the Fourier
momentum ${\bf k}$ is inside the region $\Omega(t)$ defined by the condition 
\begin{equation}
G^{\mathrm P}({\bf k}\in{\Omega(t)})\geq t
\label{omega}
\end{equation}
(see figure \ref{figure1}) and according to (\ref{cutoff}) $\partial_t
G=0$ outside of $\Omega$. For the purposes of the present paper
it will be sufficient to keep in the fully renormalized quantities
only the zero-momentum component $s_{\bf k=0}$. Therefore, in the
course of the RG flow the components outside $\Omega$ can be set
equal to zero \cite{wilson}: $s_{\bf k}|_{{\bf k}\in\bar{\Omega}}=0$ where
$\bar{\Omega}=\mbox{BZ}\backslash\Omega$ is the part of 
BZ external to $\Omega$. Next one may rescale the momenta in
(\ref{ULPA}) so that $\Omega$ regained the same volume as the original
BZ and the rescaled system could be considered as describing the same
lattice model approximately renormalized by the Kadanoff real-space
procedure \cite{block-spins1966}. 

Further details on the rescaling can be found in \cite{1984} but in this
paper the RG LPA equation of \cite{1984} in the scaling form will be
used only for the calculation of critical exponents. For the calculation
of non-universal quantities, however, the rescaling has been omitted
because it introduces a large but trivial Lyapunov exponent equal to the
spatial dimension of the system which makes the RG equation numerically
unstable. Instead, we simply substitute (\ref{ULPA}) in (\ref{exact0})
and observe that the second derivative on the r.h.s.\ makes all terms in
(\ref{ULPA}) independent of ${\bf k}$ so they all acquire the common
factor \cite{sc-lpa-rg}
\begin{equation}
p(t)=\frac{1}{N}\sum_{\bf k}\theta\left[G({\bf k})
-t \right] =\int_0^{t^{-1}-r}dE \rho(E)
	\label{p}
\end{equation}
where $\rho(E)$ is the density of states corresponding to the dispersion
$\epsilon({\bf k})$ \cite{latticeRG2010,tokar2019effective,sc-lpa-rg}. 

In the second term on the l.h.s. of (\ref{exact0}), however,
the summation over ${\bf k}$ is lifted by one of the two Kronecker
deltas so the dependence on the momentum remains in the cutoff function
(\ref{cutoff}). In the LPA we neglect it by assuming that in the critical
region the most important are small momenta for which (\ref{G_t=1}) holds.
In \ref{lat-vs-continuum} it has been argued that this approximation
should be better fulfilled in the lattice case than in the continuum
Landau-Ginzburg model \cite{1984}.  Thus, the LPA RG equation for the
local potential (\ref{u(xt)}) in the Ising universality class reads
\begin{equation}
       u_t + \frac{1}{2}u_x^2
=\frac{1}{2}p(t)u_{xx}.
       \label{LPA}
\end{equation}
To make notation more compact, here we have made use of the fact that in
the Ising universality class in the LPA there is no need for the vector
component subscripts so henceforth they will be used to denote partial
derivatives. In this notation the free energy (\ref{FH}) per site in
LPA reads
\begin{equation}
f(h)= u^{\mathrm R}(h/r)-h^2/2r
-\frac{1}{2N}\Tr\ln{\left(2\pi\hat{{G}}^{\mathrm P}\right)}
\label{f}
\end{equation}
where use has been made of (\ref{ZR}) and the argument of $u^{\mathrm
R}$ has been found by substituting (\ref{s0}) in (\ref{ULPA}) at
the end of renormalization when only ${\bf k=0}$ component contributes
to $U^{\mathrm R}$ which means that at $t^{\mathrm R}$ the argument of
$u^{\mathrm R}$ is
\begin{equation}
x=h/r.
\label{xR}
\end{equation}
The equation of state is obtained as
\begin{equation}
m(h)=-\frac{df}{dh}=\frac{h}{r}
-\frac{1}{r}u^{\mathrm R}_x\vert_{x=h/r}
\label{eq-of-state}
\end{equation}
and the SC condition (\ref{SCU}) in the LPA takes the form
\begin{equation}
\left.u^{\mathrm R}_{xx}\right\vert_{x=h/r}=0.
\label{SCu}
\end{equation}

The general $n$-vector case can be treated similarly \cite{1984}
and in fully $O(n)$-symmetric case the LPA equation simplifies to
\cite{tokar2019effective}
\begin{equation}
u_t + \frac{1}{2}(\nabla u)^2=\frac{1}{2}p(t)\nabla^2 u
\label{LPAn}
\end{equation}
where the differential operators act in the space of $n$-vectors ${\bf x}$.
\section{\label{illustrations}Illustrative calculations}
\subsection{\label{phi4}Critical temperatures in $\varphi^4$ model}
The first example we consider is the $\varphi^4$ model on sc lattice
\cite{hasenbusch_phi4,CAILLOL2012291}.  In notation of section \ref{defs}
the interaction potential reads
\begin{equation}
U_0(\vec{s})= \sum_i\left[(1-3K-r/2)s_i^2+\lambda(s_i^2-1)^2\right]
\label{Uphi4}
\end{equation}
(cf.\ equation (2) in \cite{hasenbusch_phi4}). The LPA solution has
been obtained as follows. First $U_0$ in (\ref{Uphi4}) has been mapped
onto $u(x,0)$ by replacing $s_i$ with $x$ in the summand. Then 
equation (\ref{LPA}) has been solved numerically for $h=0$ (the critical
value of the field) and iterated to satisfy the SC condition (\ref{SCu})
(for details of the procedure see \cite{sc-lpa-rg}) to determine $r$
corresponding to chosen parameters $K$ and $\lambda$. The critical value
$K_c$ corresponds to $r=0$. Calculations within the method of lines have
been performed for 11 values of $\lambda$ in the range 0.1\textemdash2.5
that were used in MC simulations in \cite{hasenbusch_phi4}.  The
calculated $K_c$ agreed with the MC values with accuracy better than
0.25\% similar to the LPA calculations in \cite{CAILLOL2012291} based
on EA EFRGE \cite{wetterich_exact_1993,morris1994,berges_non-perturbative_2002,%
dupuis_nonperturbative_2021} and the step function cutoff. Moreover,
even the systematic change with $\lambda$ of the discrepancy sign 
has been similar in both calculations, presumably, as a consequence of
equivalence of the exact equations and of the same cutoff used in the
LPA in \cite{CAILLOL2012291} and in the present paper.
\subsection{\label{spin-lattice}LPA for spin-lattice models}
The $O(n)$-symmetric spin-lattice models in principle can be solved in
the same way as the $\varphi^4$ model, only using equations (\ref{LPAn})
instead of (\ref{LPA}). To this end it would be sufficient to replace in
(\ref{Uphi4}) $s_i$ by $n$-dimensional vectors ${\bf s}_i$ and to let
$\lambda\to\infty$ to suppress fluctuations of the vector length. However,
numerical solution of (\ref{LPAn}) with infinite initial condition would
be technically very difficult. But in the partition function (\ref{Z})
the $\lambda\to\infty$ limit amounts to appearance in the integrand
of the factor $\prod_i\delta({\bf s}_i^2-1)$ \cite{n-vector-models}
which is site-local and despite being singular it can be exactly
renormalized with the use of (\ref{f_ini}) which has been generalized
to the $O(n)$ case in \ref{ini}.  The exact renormalization has been
done in the interval (see figure \ref{figure1})
\begin{equation}
0\leq t\leq t_0=\min_{\bf k}G^{\mathrm P}({\bf k})
=\frac{1}{r+\max_{\bf k}\epsilon({\bf k})}
\label{t0}
\end{equation}
where (\ref{G_t=1}) is satisfied exactly because the integral in
(\ref{p}) saturates to unity above the upper limit of the dispersion.
Explicit expressions for $u^{(n)}({\bf x},t_0)$ from (\ref{n=2-5}) can
be used as the initial conditions for LPA RG equation (\ref{LPAn}). In
the symmetric phase the equation depends only on the radial coordinate
and its numerical integration for $n>1$ has been as straightforward as
in $n=1$ case \cite{tokar2019effective}. The critical temperatures in
table \ref{T_c} have been found as the points where the SC values of $r$
interpolated to zero. The accuracy similar to that in table \ref{T_c}
was also obtained in RG calculations for IM \cite{parola_recent_2012}
and additionally for $XY$- and the Heisenberg models on the cubic lattice
in \cite{latticeRG2010}. In the later case, however, it was found to
be necessary to resort to a non-functional technique and to introduce
heuristic modifications in the partially renormalized EA which makes
the good accuracy obtained less convincing than in the present approach.
\begin{table}
\caption{\label{T_c}Dimensionless inverse critical temperatures of the
$n$-vector spin models on cubic lattices calculated in the LPA. The
errors have been estimated by comparison with the MC simulations data
\cite{bcc-fcc-diamond-Kc} for $n=1$ (IM) and with the high
temperature expansion for $n=2$, 3 and 4 \cite{n-vector-models}.}
\begin{tabular}{@{}llcl}
\hline
$n$&Lattice&$K_c$&Error\\
\hline
1&fcc&0.1023&0.2\%\\
1&bcc&0.1579&0.3\%\\
1&sc &0.2235&0.8\%\\
2&bcc&0.3225&0.6\%\\
2&sc &0.4597&1.2\%\\
3&bcc&0.4905&0.8\%\\
3&sc &0.7025&1.4\%\\
4&bcc&0.6608&0.8\%\\
4&sc &0.9488&1.4\%\\
\hline
\end{tabular}
\end{table}
As is seen, the values of critical temperatures have been calculated
with the accuracy better than 2\% which should be sufficient for most
practical purposes because in realistic lattice models microscopic
Hamiltonians are rarely known with better accuracy.
\subsection{\label{exponents}Critical exponents}
The critical exponents have been calculated with the use of the scaling
form of (\ref{LPAn}) \cite{1984}. Because in the LPA there is only
one independent exponent, for comparison purposes exponent $\nu$ has
been used. The values found were 0.65(0.63) for $n=1$, 0.71(0.67) for
$n=2$, 0.76(0.71) for $n=3$ and 0.80(0.75) for $n=4$; in parentheses
are shown the rounded values from precise calculations taken from
\cite{RG2002review}. The LPA values were closest to those calculated in
\cite{berges_non-perturbative_2002} in the lowest order of the derivative
expansion, though being systematically larger on $\sim0.01$. This
apparently is a consequence of a similar non-perturbative RG approach
used by the authors while the difference could be attributed to the fact
that $\eta$ in the calculations in the above reference was not equal to
zero but obtained from RG equations.

Though not very large, the errors in the LPA values of critical exponents
can be seen in the discrepancy between experimental and theoretical curves
in the disordered phase in \cite{sc-lpa-rg} and can be even larger in
other experiments, e.g., on the specific heat where exponent $\alpha$
is small and the LPA error is about 100\% in the Ising universality class.

Going beyond LPA within lattice nonperturbative RG, however,
is not an easy task mainly because of the large values of the
momenta within BZ which precludes the use of DE; seemingly more
appropriate expansion in circular harmonics, on the other hand,
worsens the accuracy of critical temperatures which is undesirable
\cite{dupuis_non-perturbative_2008,latticeRG2010}. From a practical
standpoint it seems reasonable to exploit the fact that in the critical
region the interactions in the course of the RG flow diminish to the point
where the perturbative treatment becomes justifiable \cite{wilson}.
This makes possible to invoke the multi-step renormalization technique to
switch from LPA to a perturbative treatment in a consistent manner at
an appropriately chosen late-stage point $t_1\lesssim t^{\mathrm R}$
(see figure \ref{figure1}).  Using the Hori representation (\ref{hori})
the usual diagrammatic techniques can be used in the calculations with the
only modification that the conventional propagator that behaves at small
$k$ as $G^{\mathrm P}(k)\simeq 1/(ck^2+r)$ (see (\ref{k20})) should be
replaced according to (\ref{Delta}) by
\begin{equation}
\Delta(k,t^{\mathrm R},t_1)\simeq \frac{1}{ck^2+r} -\frac{1}{c\Lambda^2+r}
\label{Gapprox}
\end{equation}
where $\Lambda$ is the momentum cutoff corresponding to $t_1$ (see
discussion after equation (\ref{quadratic_term})). As is seen, when
$r,k\to0$ the conventional first term in (\ref{Gapprox}) dominates
so the leading terms in the perturbative expansion will coincide
with the standard theory. The diagrams containing the second term
will differ from the leading contribution in that the propagator
will be replaced by a constant and so the diagram should be easier
to calculate than the main contribution. It may be hoped that when
$t_1$ is sufficiently close to $t^{\mathrm R}$ the corrections
to the critical temperature will be small and good precision of
$K_c$ can be preserved. The critical exponents in this approach
will be the same as those calculated within the chosen perturbative
technique \cite{wilson,Pelissetto2002CriticalPA,PhysRevE.101.042113,%
shalaby_critical_2021}.
\section{\label{fopts}First order phase transitions}
For simplicity we restrict our discussion of FOPTs to the ferromagnetic IM
where transitions occur below the critical temperature when the external
magnetic field $h$ changes between infinitesimally small values $0^-$ and
$0^+$. During the change the spontaneous magnetization jumps from $-m_0$
to $m_0>0$ or vice versa so the susceptibility $dm/dh$ is infinite inside
the coexistence region but in the immediate vicinity of $h=0$ it should
be finite. This behaviour means that the free energy per site $f(h)$
contains a singular contribution of the form $f_{sing}\simeq m_0|h|$
\cite{nienhuis_first-order_1975,fisher_scaling_1982} with the respective
term in the local potential 
\begin{equation}
u_{sing}^{\mathrm R}=-rm_0|x|. 
\label{us}
\end{equation}
Though mathematically simple, this term is not straightforward to obtain
as a solution of RG equations because the local potential (\ref{u(xt)})
is assumed to be an analytic function of the field variable.

The origin of the singularity can be elucidated by differentiating
LPA equation (\ref{LPA}) w.r.t.\ $x$ to arrive at the generalized BE
\cite{gbe}
\begin{equation}
	\mu_t+\mu\mu_x=(p/2)\mu_{xx},	
	\label{burgers}
\end{equation}
where
\begin{equation}
	\mu(x,t)=u_x(x,t).
	\label{mu}
\end{equation}
According to (\ref{us}), $\mu^{\mathrm R}(x)$ should be discontinuous at
$x=0$.  As is known \cite{whitham_linear_2011}, the discontinuous shock
waves appear in the inviscid BE which in our case would mean vanishing
$p$.  According to (\ref{p}), $p=0$ at the end of renormalization when
$t^{\mathrm R}=1/r$. It is important to note that in models studied in
\cite{sc-lpa-rg,tokar21BC} and in the present paper $p(t)$ additionally
satisfied the condition $(dp/dt)_{t^{\mathrm R}}=0$ due to $\rho(E=0)=0$
in (\ref{p}).  It seems that the latter condition should also be satisfied
for the presence of FOPTs because the calculations with a model which
did not satisfy it the discontinuous solutions could not be found.

To describe FOPTs with the use of the equation of state
(\ref{eq-of-state}) it would be sufficient to solve (\ref{burgers})
and (\ref{mu}) using the method of lines \cite{Mousa+2015+47+58}.
However, because $\mu^{\mathrm R}(x)$ is discontinuous, its first
and the second derivatives entering (\ref{burgers}) will be difficult
to deal with numerically.  Therefore, in the actual calculations in
\cite{sc-lpa-rg,tokar21BC} the difficulties were alleviated by using the
integrated form of (\ref{burgers}), that is, the original LPA equation
(\ref{LPA}). Furthermore, additional regularization has been found
to be useful that can be achieved with the use of the partial LT 
(\ref{y-x1})--(\ref{v-u1}). By solving the transformed equation
(\ref{LPA2}) instead of (\ref{LPA}) it can be seen from (\ref{y-x1}) that
$x=h/r$ at $t^{\mathrm R}$ so the jump in $u_x$ at $x=0$ is transformed
in a finite interval of $y$ values. Thus, the only singularities in
$v^{\mathrm R}(y)$ will be two kinks at points $y^\pm_0=-\bar{t}^{\mathrm
R}u^{\mathrm R}_x\vert_{x=0^\pm}$ while inside the interval $\mu^{\mathrm
R}(y)=v^{\mathrm R}_y(y)$ will be a linear function of $y$, as can be seen
from (\ref{h-y}) for $h=0$. So the strongest singularities in equation
(\ref{LPA2}) will be two discontinuities in the second derivative at
the kink points corresponding to the jump of the susceptibility at the
boundary separating the ordered phase and the coexistence region.

The mechanism of appearance of the linear in $y$ segment in $\mu(y,t)$ can
be qualitatively understood from the equation obtained by differentiating
(\ref{LPA2}) w.r.t.\ $y$:
\begin{equation}
	\mu_t = \frac{p(t)\mu_{yy}}{2(1+\bar{t} \mu_{y})^2}.
	\label{LPA21}
\end{equation}
As is seen, when the term in parentheses vanishes the r.h.s.\
becomes singular.  Because $\bar{t}\geq0$ this may happen only
for negative values of $\mu_{y}$ which appear when $v(y,t)$ has a
negative curvature in some region of $y$ values. The latter appears
in the initial potentials (\ref{n=2-5}) $u$ or $v$ which are equal
due to (\ref{v-u1}) at sufficiently low temperatures.  Further, the
coefficient before $\mu_{yy}$ in (\ref{LPA21}) is always positive
and so can be considered as a space and time dependent diffusivity.
At vanishing denominator the singularity becomes non-integrable
so the maximum negative slope tolerated by equation (\ref{LPA21})
is $-1/\bar{t}^{\mathrm R}$; steeper slopes will be smeared out by
the diffusion because the diffusivity will diverge in such cases. In
numerical calculations in \cite{tokar2019effective,sc-lpa-rg,tokar21BC}
and in the exactly solvable model below it was found that the fully
renormalized solution exhibited the universal behaviour $\mu^{\mathrm
R}(y)=-y/\bar{t}^{\mathrm R}+(f.i.t.)$ corresponding to the maximum
allowable steepness.  Substitution in (\ref{h-y}) shows that the change
of $\mu^{\mathrm R}(x)$ within the coexistence region is confined to a
single point $x=0$, as expected.

The above qualitative reasoning can be illustrated with an exactly
solvable example provided by the infinite-range IM (IRIM) also known
as the Husimi-Temperley \cite{Brankov_1983} and the Curie-Weiss model
\cite{salinas_introduction_2001} which is often considered to be exactly
solvable in the MF approximation \cite{salinas_introduction_2001}. A
straightforward application of the MF theory, however, leads to unphysical
van der Waals loops in the FOPT region which suggests that the MF solution
below $T_c$ is flawed.  In \cite{Brankov_1983,choquard_mean_2004} it was
shown that the loops are replaced by the shock wave solutions of the BE
that can be derived for that particular model without resort to RG.

In \ref{irim} it has been shown that exactly the same BE as in
\cite{Brankov_1983,choquard_mean_2004} is obtained for this model
as the S-matrix RG equation in the LPA. In the IRIM case $p(t)$
in (\ref{LPA}) becomes equal to $1/N$ so that the generalized BE
(\ref{burgers}) turns into the conventional BE with constant viscosity
$1/2N$. To our purposes the most important is the fact that the jumps in
$\mu^{\mathrm R}(x)$ as the external field crosses zero are between the
MF spontaneous magnetisation values $\pm m_0$. Thus, though $\mu=u_x$ in
(\ref{burgers}) and $m(x)$ in \cite{Brankov_1983} do not coincide, from
(\ref{eq-of-state}) it can be seen that at $h=0$ the functions in both
cases are the same. Now taking into account that our LPA RG solution in
\ref{irim} coincides with the MF solution \cite{salinas_introduction_2001}
outside the coexistence region and unifying it with the rigorous treatment
in \cite{Brankov_1983} of discontinuities in the solution of BE which is
equally valid for our $h=0$ case we conclude that the LPA RG equation
solves exactly both the IRIM and the problem of the van der Waals
loops. Finally, in order to check whether the numerical solution by the
method of lines used in \cite{tokar2019effective,sc-lpa-rg,tokar21BC}
gives in the IRIM case the same solution as the analytic approach,
(\ref{LPA2}) has been solved with $p=1/2N$ for $N=1000$ and the exact
solution has been reproduced with the accuracy $\Or(10^{-3})$.

In connection with the numerical solutions it is pertinent to note that
though (\ref{LPA2}) in the FOPT region has been found to be easier to deal
with than with equation (\ref{LPA}) \cite{sc-lpa-rg,tokar21BC}, it may
still be worthwhile to adopt the techniques of \cite{Mousa+2015+47+58}
to the case of (\ref{LPA}).  The reason is that the phase transitions
in $u^{\mathrm R}(x)$ take place at a single value of argument $x$
while in (\ref{LPA2}) $x$ is mapped on an interval of $y$ values of
length $\Or(1)$. But $v^{\mathrm R}(y)$ within the interval behaves
linearly and in the symmetric case can be fully characterized by a
single parameter, the slope. However, for good numerical accuracy
the discretization step has to be chosen to be $\Or(10^{-3})$ or
smaller \cite{caillol_non-perturbative_2012,sc-lpa-rg}. Thus, in the
method of lines $\Or(10^3)$ points carry essentially the same physical
information. Because in practice the maximally possible number of lines
is restricted, the use of (\ref{LPA}) may prove to be more suitable for
achieving better accuracy.
\section{\label{discuss}Conclusion}
In the extensive review of the nonperturbative RG approach
\cite{dupuis_nonperturbative_2021} the authors motivated the
need for the modern implementation of Wilson's RG suggested in
\cite{wetterich_exact_1993,BONINI1993441,ellwanger1993,morris1994}
by the complexity of previously derived EFRGEs
\cite{wilson,wegner_renormalization_1973,nicoll_exact_1976} which
impeded the development of reliable approximate computational schemes.
The results of the present paper can be considered as further advance
in this direction in the case of the classical lattice models.

The new advancements include such important achievement as the
derivation of the S-matrix EFRGE and of its Legendre transformed
form which account for all contributions to the free energy and
EA functionals, not only for the field-dependent terms as is
conventional in the Wilsonian RG \cite{wilson,morris1994}. This
has made possible the RG calculation of all thermodynamic
quantities in lattice systems.  Moreover, by LT of the S-matrix
EFRGE it has been possible to identify the cause of the divergence
inherent in the conventional formulation of the Wetterich equation
\cite{wetterich_exact_1993,morris1994,berges_non-perturbative_2002,%
dupuis_nonperturbative_2021}. An explicit expression for a divergent
conterterm has been derived which having being subtracted from the fully
renormalized solution leads to the exact finite expression for the Gibbs
FE. This allows one to obtain with the use of the Wetterich equation
the same results as with the S-matrix EFRGE.

Further, within the S-matrix RG approach a thermodynamic
description of the FOPT of general kind may be achieved. By the latter
are meant the FOPT not only due to the symmetry breaking as in
the IM below $T_c$ but also those that are defined by the crossing of
the FE curves, such as in the Blume-Capel model below the tricritical
point \cite{tokar21BC}.  The description is further facilitated by the
structure of the S-matrix EFRGE, especially in the LPA, which
can be straightforwardly connected with the generalized BE through its
differentiation w.r.t.\ the field variable. This has made possible to
generalize on the nonperturbative RG the observation made in connection
with the MF solution of IRIM in \cite{Brankov_1983,choquard_mean_2004}
that FOPT correspond to the shock-wave solutions of RG equations.  In the
case of EA EFRGEs this could be approximately justified only in the
large-$n$ limit \cite{grossi2019resolving} which excludes the majority of
physically important small-$n$ models, such as the IM, the Blume-Capel
model, etc.. The connection of S-matrix EFRGE with the generalized
Burgers' equation has made possible to apply in the solution of RG
equations the concepts and techniques developed in the field of nonlinear
differential equations \cite{whitham_linear_2011,gbe,gao_analytical_2017}.

Some useful improvements in the S-matrix approach to the lattice models 
in comparison with the approach of \cite{latticeRG2010,CAILLOL2012291}
can be mentioned.  For example, the problem of establishing the initial
condition does not arise in the S-matrix case; the need to calculate
improper integrals is absent because the momenta much larger than the
inverse lattice spacing do not appear in the formalism. All necessary
LTs have been found in explicit analytic form and the need for their
numerical calculation do not arise. These improvements are important in
the solution of the evolution equations of RG type because the positive
Lyapunov exponents inherent in RG in the critical region may enhance
small errors in the initial condition resulting in degraded accuracy.

The absence of large unphysical momenta in the S-matrix EFRGE is
also important from the point of view of qualitative soundness of the
approximate solutions obtained. In the absence of rigorous methods of
solution of strongly coupled field-theoretic models, qualitative arguments
play a significant role. Therefore, the use in the S-matrix approach
of the SC condition and of the amputated correlation functions in the
renormalized interaction potential makes LPA a natural approximation from
a qualitative standpoint.  Furthermore, the use of the cutoff in the form
of the step function \cite{1984,latticeRG2010} in the spin-lattice case
leads to the exact renormalization at the earliest stage of the RG flow
when the Fourier momenta are not yet eliminated.  In the present paper
it has been argued that at an early stage of the momenta elimination
the gradual loss of LPA accuracy caused by the widening gaps in the
periodic zone scheme is accompanied by the weakening of interactions. So
that at the late stage of the RG flow they may become sufficiently weak
for LPA to be justifiable by DE \cite{Morris_1999}.  This qualitative
reasoning agrees with the results of \cite{CAILLOL2012291} where the
use of the step function cutoff in the strong coupling case resulted in
considerably more accurate values of critical temperatures than with the
Wilson-type cutoff. The values of the critical temperatures calculated
in \cite{CAILLOL2012291} within the EA LPA in $\varphi^4$ model are very
similar to those of the present paper based on the S-matrix equation.
Apparently this is a consequence of the equivalence of the EA and the S-matrix
equations also in the LPA. The high accuracy of the obtained results
further supports the qualitative correctness of the picture of the
RG flow described by the S-matrix EFRGE.  

Further, the possibility to easily implement a multi-step renormalization
within the S-matrix formalism makes possible to use in the critical region
the accurate perturbative results obtained, for example, with the use
of DE \cite{PhysRevE.101.042113,shalaby_critical_2021}. Such an approach
may be useful in correcting the deviations of the LPA predictions from
the experimental data caused by errors in the LPA critical exponents
\cite{sc-lpa-rg}.

To sum up, the main advantage of the RG approach based on the S-matrix
EFRGE is that it provides conceptually simpler and computationally easier
techniques of solution of spin-lattice models than the alternative
methods \cite{latticeRG2010,parola_recent_2012} while preserving
the same accuracy  in the calculation of the critical temperatures
as was achieved in \cite{latticeRG2010,CAILLOL2012291}. Besides, the
possibility to calculate correct values of the f.i.t.\ in FE within the
S-matrix formalism allows for its use in the description of FOPT of
any kind \cite{tokar21BC}.
\ack
This research was supported by the National Academy of Sciences of
Ukraine under contract No. 22/20-H. I am particularly indebted to Hugues
Dreyss\'e for his support and encouragement. I expresses my gratitude
to Universit\'e de Strasbourg and IPCMS for their hospitality.
\appendix 
\section{\label{steps}Stepwise renormalization in the Green's function formalism}
Renormalization in several finite steps can be performed recursively by
adding steps one at a time so only a two-step case need be considered.
This can be done with the use of the identity satisfied by the Gaussian
kernel
\begin{eqnarray}
\int_{-\infty}^\infty d\vec{s}_0\;
{\det\left(\frac{\hat{X}}{2\pi}\right)}^{\frac{1}{2}}
\rme^{-\frac{1}{2}(\vec{s}-\vec{s}_0)^\dagger\hat{X}(\vec{s}-\vec{s}_0)}
\;{\det\left(\frac{\hat{Y}}{2\pi}\right)}^{\frac{1}{2}}
\rme^{-\frac{1}{2}(\vec{s}_0-\vec{s}\,^\prime)^\dagger\hat{Y}(\vec{s}_0-\vec{s}\,^\prime)}\nonumber\\
={\det\left(\frac{\hat{X}\hat{Y}}{2\pi(\hat{X}+\hat{Y})}\right)}^{\frac{1}{2}}
\rme^{-\frac{1}{2}(\vec{s}-\vec{s}\,^\prime)^\dagger
\frac{\hat{X}\hat{Y}}{(\hat{X}+\hat{Y})}(\vec{s}-\vec{s}\,^\prime)}
\label{group}
\end{eqnarray}
where all matrices are considered to be translationally-invariant, hence,
commuting with each other.

Now by assuming that the second kernel on the first line is
${\cal G}(\vec{s}_0,\vec{s}\,^\prime, t_0)$ from (\ref{green}) with
$\hat{Y}=\hat{R}(t_0)$ and that on the second line we want to obtain
(\ref{green}), we conclude that
\begin{equation}
\frac{\hat{X}\hat{R}(t_0)}{\hat{X}+\hat{R}(t_0)}=\hat{R}(t).
\label{XYR}
\end{equation}
Solving this one finds that the regulator matrix for the second step is 
\begin{equation}
\hat{X}(t,t_0)=\left[\hat{R}^{-1}(t)-\hat{R}^{-1}(t_0)\right]^{-1}=
\left[\hat{G}(t)-\hat{G}(t_0)\right]^{-1}
\label{X}
\end{equation}
As is seen, $\hat{X}$ satisfies the initial condition for Green's
functions $\hat{X}(t\to t_0,t_0)\to\infty$; besides, because
$\hat{R}(0)=\infty$, $\hat{X}(t,0)=\hat{R}(t)$ so (\ref{X}) is valid
for all $t$ and $t_0\leq t$.
\section{\label{legendre}Legendre transforms}
By analogy with the LTs of
\cite{x-yLegendre,tokar2019effective} we introduce a new fluctuating
field $\vec{\phi}$ and a new interaction potential $W$ as
\begin{eqnarray}
	\label{y-s-k}
\phi_{\bf k}({\vec{s}},t)&=&s_{\bf k}-G({\bf k},t) \frac{\partial
U({\vec{s}},t)}{\partial {s_{\bf -k}}}\\
	\label{W-U-k} W(\vec{\phi},t)&=&U({\vec{s}},t)
-\frac{1}{2} \sum_{\bf q}\frac{\partial U({\vec{s}},t)}{\partial {s_{\bf -q}}}
G({\bf q},t)\frac{\partial U({\vec{s}},t)}{\partial {s_{\bf q}}} 
\end{eqnarray}
where we assume $G({\bf k},t)$ to be a non-negative even function of
${\bf k}$ satisfying $G({\bf k},0)=0$.

To transform the exact RG equation (\ref{exact0}) to the new variables
we first take the derivatives of (\ref{y-s-k}) and (\ref{W-U-k})
w.r.t. $s_{{\bf k}^\prime}$ 
\begin{eqnarray}
	\label{diff-y-s-k}
&&\frac{\partial \phi_{\bf k}}{\partial s_{{\bf k}^\prime}} =\delta_{{\bf
k}{\bf k}^\prime}-G({\bf k},t) \frac{\partial^2 U}{\partial {s_{\bf
-k}}\partial s_{{\bf k}^\prime}}\\
	\label{diff-W-U-k}
&&	\sum_{\bf k}\frac{\partial W}{\partial {\phi_{\bf k}}}
\frac{\partial \phi_{\bf k}}{\partial s_{{\bf k}^\prime}} =\frac{\partial
U}{\partial {s_{{ \bf k}^\prime}}} -\sum_{\bf q}\frac{\partial
U({\vec{s}},t)}{\partial {s_{\bf q}}}G({\bf q},t) \frac{\partial^2 U}{\partial
{s_{\bf -q}}\partial s_{{\bf k}^\prime}} 
\end{eqnarray} 
and observe that by substituting (\ref{diff-y-s-k}) in (\ref{diff-W-U-k})
and changing the summation subscript from ${\bf k}$ to ${\bf q}$
one arrives at the system of linear equations for the row vector
$\left({\partial W}/{\partial {\phi_{\bf q}}} -{\partial U}/{\partial
{s_{\bf q}}}\right)^\dagger$
\begin{equation} 
\sum_{\bf q} \left(\frac{\partial W}{\partial {\phi_{\bf
q}}} -\frac{\partial U}{\partial {s_{\bf q}}}\right)^\dagger \left(\delta_{{\bf
q}{\bf k}^\prime}-G({\bf q},t) \frac{\partial^2 U}{\partial {s_{\bf
-q}}\partial s_{{\bf k}^\prime}} \right)=0.  
\label{identity1}
\end{equation} 
Because $G({\bf q},t)$ can be arbitrary, the matrix of
the system in general is not degenerate which means that the solution
should be trivial and the following useful equality should hold 
\begin{equation}
\frac{\partial W}{\partial {\phi_{\bf q}}} =\frac{\partial U}{\partial
{s_{\bf q}}}.  
\label{V_y=U_s} 
\end{equation} 
In particular, with its use transformations (\ref{y-s-k}) and
(\ref{W-U-k}) are easily reversed. For example, from (\ref{y-s-k})
one gets
\begin{equation}
s_{\bf k}(\vec{\phi},t)=\phi_{\bf k}+G({\bf k},t)\frac{\partial W} {\partial
{\phi_{\bf -k}}}(\vec{\phi},t) 
\label{s-y-k} 
\end{equation}

The next step in transforming the RG equation is to differentiate
(\ref{y-s-k}) and (\ref{W-U-k}) w.r.t.\ $t$: 
\begin{eqnarray}
	\label{dt-y-s-k}
 &&\partial_t \phi_{{\bf k}}=
-\partial_t G({\bf k},t)\frac{\partial U}{\partial {s_{\bf -k}}} -G({\bf
k},t) \frac{\partial^2 U}{\partial {s_{\bf -k}}\partial t }\\
	\label{dt-W-U-k}
&&\partial_t W+ \sum_{\bf k}\frac{\partial W}{\partial {\phi_{\bf
k}}} \partial_t\phi_{{\bf k}}=\partial_tU -\sum_{\bf q}\frac{\partial
U}{\partial{s_{\bf q}}} G({\bf q},t)\frac{\partial^2 U}{\partial {s_{\bf
-q}}\partial t} \nonumber\\ &&-\frac{1}{2} \sum_{\bf q}\frac{\partial
U} {\partial{s_{\bf -q}}}\partial_tG({\bf q},t)\frac{\partial U}
{\partial{s_{\bf q}}} 
\end{eqnarray} 
By substituting (\ref{dt-y-s-k})
in (\ref{dt-W-U-k}) and using (\ref{V_y=U_s}) one can express the
terms on the l.h.s.\ in (\ref{exact0}) in terms of $W(\vec{\phi},t)$ as
\begin{equation} \partial_tU+\frac{1}{2} \sum_{\bf q}\frac{\partial
U} {\partial{s_{\bf -q}}}\partial_tG({\bf q},t)\frac{\partial U}
{\partial{s_{\bf q}}}=\partial_t W.  \label{2terms} \end{equation}
The r.h.s.\ in (\ref{exact0}) can be transformed by differentiating
(\ref{V_y=U_s}) and (\ref{s-y-k}) w.r.t.  $\phi_{{\bf -k}^\prime}$ as
\begin{eqnarray} 
&&\sum_{\bf q}\frac{\partial^2 U}{\partial s_{{\bf
k}} \partial {s_{\bf -q}}}\left[\delta_{{\bf q},{\bf k}^\prime}
+G({\bf q},t)\frac{\partial^2W}{\partial{\phi_{\bf q}\partial
\phi_{{\bf -k}^\prime}}}\right] =\frac{\partial^2W}{\partial{\phi_{\bf
k}\partial \phi_{{\bf -k}^\prime}}} 
\label{V_yy-U_ss} 
\end{eqnarray} 
Now denoting the matrix in the brackets by $\hat{B}=[B_{\bf q,k^\prime}]$
(\ref{V_yy-U_ss}) can be solved as
\begin{equation} \frac{\partial^2
U}{\partial s_{{\bf k}} \partial {s_{\bf -k}}}=\sum_{\bf k^\prime}
\frac{\partial^2W}{\partial{\phi_{\bf k}\partial \phi_{{\bf -k}^\prime}}}
\left(\hat{B}^{-1}\right)_{\bf k^\prime,k}.  
\label{U''proptoW''}
\end{equation} 
This expression can be substituted in (\ref{exact0})
to give the exact RG equation in terms of
the Legendre-transformed quantities $\vec{\phi}$ and $W$ 
\begin{equation}
\partial_t W = \sum_{\bf k^\prime,k}\partial_tG({\bf k},t)
\frac{\partial^2W}{\partial{\phi_{\bf k}\partial \phi_{{\bf
-k}^\prime}}} \left(\hat{B}^{-1}\right)_{\bf k,k^\prime}
\label{exact1} 
\end{equation} 
which structure is very similar (but not identical) to that of equation 
(3.31) in \cite{morris1994}.
\subsection{\label{wetterich}The Wetterich equation}
The EA EFRGE derived in \cite{wetterich_exact_1993,berges_non-perturbative_2002}
(the Wetterich equation) reads  
\begin{equation}
\partial_t\tilde{\Gamma}(\vec{\varphi},t)=\frac{1}{2}\Tr\left\{\partial_t
\hat{{R}}^{(0)} \left(\hat{\tilde{\Gamma}}_{\varphi\varphi}
+\hat{{R}}^{(0)}\right)^{-1}\right\}
\label{wetterich1} 
\end{equation} 
where $\hat{{R}}^{(0)}$ in addition to the requirement 
to all regulators (\ref{R2infty}) also satisfies the condition
\begin{equation}
\hat{ R}^{(0)}(t=t^{\mathrm R})=0
\label{R20}
\end{equation}
(the superscript (0) is used to distinguish such regulators). The RG
flow in (\ref{wetterich1}) is initiated by
\begin{equation}
\tilde{\Gamma}(\vec{\varphi},t=0)=H_0(\vec{\varphi}).
\label{gamma-ini}
\end{equation}

To establish its connection with (\ref{exact1}) we first derive an
expression for the Gibbs FE that is equal to the fully renormalized EA. To
this end using Fourier transformed (\ref{s_av}) and (\ref{FH}) one finds
\begin{equation}
	\label{y-s-kR}
m_{\bf k}=G^{\mathrm P}({\bf k})h_{\bf k}-G^{\mathrm P}({\bf k})
\left.\frac{\partial U^{\mathrm R}({\vec{s}})}{\partial {s_{\bf -k}}}
\right|_{s_{\bf k}=G^{\mathrm P}({\bf k})h_{\bf k}}.  
\end{equation}
Solving this w.r.t.\ the derivative 
\begin{equation}
\left.\frac{\partial U^{\mathrm R}({\vec{s}})}{\partial {s_{\bf -k}}}
\right|_{s_{\bf k}=G^{\mathrm P}({\bf k})h_{\bf k}}  =
h_{\bf k}-\left[G^{\mathrm P}({\bf k})\right]^{-1}m_{\bf k},
\label{dUds}
\end{equation}
substituting it in (\ref{W-U-k}) at $t^{\mathrm R}$ where $\phi_{\bf
k}=m_{\bf k}$ and using (\ref{GP}) one gets in the vector notation
\begin{equation}
W^{\mathrm R}(\vec{m})+\frac{1}{2}
\vec{m}\,^\dagger(\hat{\epsilon} +r\hat{I})\vec{m}
=U^{\mathrm R}(\hat{G}^{\mathrm P}\vec{h})
-\frac{1}{2}\vec{h}^\dagger\hat{{G}}^{\mathrm P}\vec{h}
+\vec{m}\,^\dagger\vec{h}.
\label{WRUR}
\end{equation}
Comparing this with (\ref{FH}) and adding the last term in (\ref{FH})
to both sides 
\begin{equation}
W^{\mathrm R}(\vec{m})+\frac{1}{2}
\vec{m}\,^\dagger(\hat{\epsilon} +r\hat{I})\vec{m}
-\frac{1}{2}\Tr\ln{\left(2\pi\hat{ G}^{\mathrm P}\right)} 
=F^{\mathrm R}(\vec{h}) +\vec{m}\,^\dagger\vec{h}
\label{G-F}
\end{equation}
one sees that the r.h.s.\ is the $\vec{h}-\vec{m}$ Legendre transform
of the Helmholtz FE, so, by definition, it is equal to the Gibbs FE
which we will denote $\Gamma^{\mathrm R}(\vec{m})$. The expression is
naturally generalized at arbitrary $t$ as
\begin{equation} 
\Gamma(\vec{\phi},t)=W(\vec{\phi},t)+\frac{1}{2}
\vec{\phi}\,^\dagger(\hat{\epsilon} +r\hat{I})\vec{\phi}
-\frac{1}{2}\Tr\ln{\left(2\pi\hat{ G}^{\mathrm P}\right)}. 
\label{Gamma}
\end{equation} 
Further, because at $t=0$ the LT (\ref{y-s-k})--(\ref{W-U-k})
is trivial, the initial condition for (\ref{Gamma}) is obtained
by simply changing the variable from $\vec{s}$ to $\vec{\phi}$ 
so taking into account (\ref{H0}) one gets
\begin{equation}
\Gamma(\vec{\phi},t=0)= H_0(\vec{\phi})
-\frac{1}{2}\Tr\ln{\left(2\pi\hat{ G}^{\mathrm P}\right)} 
\label{Gamma0} 
\end{equation} 
which differs from the condition (\ref{gamma-ini}) of EA approach by the
second term. Because the latter is constant in $\vec{\phi}$ and $t$, it
remains unaltered in the course of the RG flow. Therefore, the constant
term instead of the initial condition (\ref{Gamma0}) can be added to
the solution at the end of the flow which below will be used in the
renormalization of $\tilde{\Gamma}^{\mathrm R}$ in the Wetterich equation.

The EFRGE for $\Gamma$ can be obtained as
follows. First, in (\ref{exact1}) we represent matrix $\hat{B}$ as the product
$\hat{G}\hat{A}$ with $\hat{G}=[G({\bf k},t)\delta_{\bf k,k^\prime}]$ and
\begin{equation} \hat{A}=\left[W_{\phi_{\bf
k}\phi_{\bf -k^\prime}} +{R}({\bf k},t)\delta_{\bf
kk^\prime}\right]\equiv\hat{W}_{\phi\phi}+\hat{{R}}. \label{A} 
\end{equation}
We remind that according to (\ref{Gdef}) $\hat{R}=\hat{G}^{-1}$,
hence, in (\ref{exact1}) $\partial_t \hat{G}=-\hat{R}^{-1}\partial_t
\hat{R}\hat{R}^{-1}$. Next, in the matrix identity
\begin{equation}
\hat{W}_{\phi\phi}\hat{B}^{-1}=(\hat{A}-\hat{R})\hat{A}^{-1}\hat{R}
=\hat{R}-\hat{R}\hat{A}^{-1}\hat{R}.  \label{identity} 
\end{equation}
the regulator function can be represented in the form
\begin{equation} 
\hat{R}=\hat{\epsilon}+r\hat{I}+\hat{{R}}^{(0)} 
\label{R-RW} 
\end{equation} 
compatible with both (\ref{RtR}) and (\ref{R20}).  Now
substituting (\ref{identity}) in (\ref{exact1}), using $t$-independence
of the last two terms in (\ref{Gamma}) and the invariance of the trace
under the cyclic permutations we arrive at an EFRGE
\begin{equation}
\fl
\partial_t\Gamma(\vec{\phi},t)=\frac{1}{2}\Tr\left\{\partial_t
\hat{R}^{(0)}
\left(\hat{\Gamma}_{\phi\phi}
+\hat{R}^{(0)}\right)^{-1}
\right\}-\frac{1}{2}\partial_t\ln\det\left(\hat{\epsilon}+r\hat{I}+\hat{{R}}^{(0)}\right)  
\label{wetterich-eq} 
\end{equation} 
where $\hat{\Gamma}_{\phi\phi}= \left[\Gamma_{\phi_{\bf k}\phi_{\bf
-k^\prime}}\right]$.

As can be seen, (\ref{wetterich-eq}) differs from the Wetterich
equation (\ref{wetterich1}) only by the last term which, however,
plays an significant role by regularizing the equation at $t\to0$
when $\hat{{R}}^{(0)}\to\infty$. Because in this limit $\Gamma$ in
(\ref{Gamma0}) is finite and so is negligible in comparison with the
regulator, the two terms on the r.h.s.\ in (\ref{wetterich-eq}) cancel to
zero at $t=0$ so unlike (\ref{wetterich1}) the equation is integrable near
$t=0$ as well as everywhere along the RG trajectory. This ensures the use
of (\ref{wetterich-eq}) in determining the exact Gibbs FE. 

Because the cancellation of singularities at $t=0$, is very important
for the consistency of (\ref{wetterich-eq}), it cannot be rearranged in
the WE form through a simple transfer of the last term on the r.h.s.\ on
the l.h.s..  To overcome this difficulty we first regularize the equation
in the conventional way by starting the evolution from small positive
value of $t=t_0$ (which is equivalent to starting from large finite
value of $\Lambda=\Lambda_0$ or $k$ in the momentum parametrization
\cite{wetterich_exact_1993,morris1994,berges_non-perturbative_2002,%
dupuis_nonperturbative_2021}). Now it is easily seen, that
(\ref{wetterich-eq}) can be cast in the form of WE (\ref{wetterich1})
for a modified EA
\begin{equation}
\fl
\tilde{\Gamma}(\vec{\phi},t,t_0)={\Gamma}(\vec{\phi},t)+
\frac{1}{2}\Tr\ln{\left(2\pi\hat{ G}^{\mathrm P}\right)} 
+\frac{1}{2}\ln\det\left.\left(\hat{\epsilon}+r\hat{I}+\hat{{R}}^{(0)}(t^\prime)\right)
\right\vert_{t^\prime=t_0}^{t^\prime=t}
\label{tilde-gamma}
\end{equation}
where the second term on the r.h.s.\ has been added in order that instead
of (\ref{Gamma0}) $\tilde{\Gamma}$ satisfied at $t_0\to0$ the initial
condition (\ref{gamma-ini}).

At the end of the evolution at $t=t^{\mathrm R}$ after some rearrangement
the fully renormalized Gibbs FE can be found with the use of the WE
solution (\ref{tilde-gamma}) as
\begin{equation}
\Gamma^{\mathrm R}(\vec{m})=\lim_{t_0\to0}
\left\{\tilde{\Gamma}(\vec{m},t^{\mathrm R},t_0)-
\frac{1}{2}\ln\det{[2\pi\hat{G}(t_0)]} \right\}
\label{renormalization}
\end{equation}
where the second term in the braces is the counterterm with $\hat{G}$
equal to the inverse of the regulator (\ref{R-RW}).

It is to be pointed out that when differentiated w.r.t.\ $\phi$
the equations (\ref{wetterich1}) and (\ref{wetterich-eq}) and their
respective initial conditions (\ref{gamma-ini}) and (\ref{Gamma0})
becomes identical which proves that both the equations and their
solutions differ only by f.i.t.. It is this differentiated form of
WE that has been used by Wetterich et al. in concrete RG calculations
\cite{ADAMS_1995,PhysRevLett.77.873}.

Thus, the Wetterich equation (\ref{wetterich1}) solved with the
conventional initial condition (\ref{gamma-ini}) will give the exact
solution of the EA EFRGE both for the field-dependent part and for
f.i.t.\ provided the divergence in the EA will be removed according to
(\ref{renormalization}).
\subsection{\label{partial}LT in a two-step renormalization} 
As could be noted, the only property of
$G({\bf k},t)$ that was used in the derivation of RG equation (\ref{exact1}) 
was that $G$ had the same time derivative
as in (\ref{exact0}).  But from (\ref{Delta}) one can see that
$\partial_t G = \partial_t\Delta$. So for the second-step renormalization
one can derive RG equation similar to (\ref{exact1}) by using the LT 
(\ref{y-s-k}) and (\ref{W-U-k}) with $G$ replaced by $\Delta$
\begin{eqnarray}
	\label{y-s-k-p}
y_{\bf k}({\vec{s}},t)&=&s_{\bf k}-\Delta({\bf k},t,t_0) \frac{\partial
U({\vec{s}},t)}{\partial {s_{\bf -k}}}\\
	\label{V-U-k-p} V(\vec{y},t)&=&U({\vec{s}},t)
-\frac{1}{2} \sum_{\bf q}\frac{\partial U({\vec{s}},t)}{\partial {s_{\bf
-q}}} \Delta({\bf q},t,t_0)\frac{\partial U({\vec{s}},t)}{\partial {s_{\bf
q}}}.  
\end{eqnarray} 
However, unless $t_0=0$, functional $V$ and variables $\vec{y}$ do
not acquire immediate physical meaning at the end of renormalization
(for $t_0=0$ they will coincide with $W$ and $\vec{\phi}$). To see
this we note that the equation for $V$ similar to (\ref{exact1}) has
as the initial condition $V(\vec{y},t_0)=U(\vec{y},t_0)$, as follows
from (\ref{y-s-k-p})--(\ref{V-U-k-p}) with $\Delta(t_0,t_0)=0$ (see
(\ref{Delta})). But according to (\ref{exact0}) $U(\vec{y},t_0)$
depends on the arbitrary values of $G({\bf k},t)$ in the interval
$0\leq t<t_0$ while in the RG equation for $V$ and in the LT 
(\ref{y-s-k-p})--(\ref{V-U-k-p}) only $G({\bf k},t)$ for
$t_0\leq t<t^{\mathrm R}$ contribute. Because of this arbitrariness,
both $V^{\mathrm R}$ and $\vec{y}$ cannot be given a physical meaning
at $t^{\mathrm R}$.  However, similar to (\ref{V_y=U_s}) the equality
$\partial V/\partial {y_{\bf q}}={\partial U}/{\partial {s_{\bf
q}}}$ holds so the transform (\ref{y-s-k-p})--(\ref{V-U-k-p}) can be
reversed and the physical quantities $U^{\mathrm R}$ and $\vec{s}$ in
(\ref{Rs=h}) can be expressed in parametric form in terms of $V^{\mathrm
R}$ and $\vec{y}$.  If needed, the Gibbs FE can be found with the use of
(\ref{G-F}).
\section{Legendre transforms in LPA}
The formalism just described considerably simplifies in the LPA.  First we
note that in the layer-cake renormalization scheme of section \ref{lpa}
visualized in figure \ref{figure1} the fluctuating field ``sees'' only
the flat part of $G({\bf k},t)=t$ (\ref{omega}) because beyond the cutoff
the field has been set to zero. Thus, $\Delta^{LPA}(t,t_0)=t-t_0=\bar{t}$
(see figure \ref{figure1}).  In terms of the local potential $v(y,t)$
corresponding to $V$ which now depends on a single scalar variable $y$
the LT (\ref{y-s-k-p})--(\ref{V-U-k-p}) simplifies to
\begin{eqnarray}
	\label{y-x1}
y&=&x-\bar{t} u_{ x}\\
	\label{v-u1}
	v&=&u-\bar{t}u_{ x}^2/2
\end{eqnarray}
where for brevity the arguments of $y(x,t)$, $v(y,t)$ and $u(x,t)$
have been omitted and the subscript notation for partial derivative has
been used.

The exact equation for $V$ is obtained from (\ref{exact1}) via
substitutions $W\to V$, $\vec{\phi}$ to $y$ and $G\to \Delta$. In the LPA it
simplifies to
\begin{equation}
	v_t = \frac{p(t)v_{yy}}{2(1+\bar{t} v_{yy})}. \label{LPA2}
\end{equation}
It is possible also to derive it directly from (\ref{LPA}) by using
(\ref{y-x1})--(\ref{v-u1}) and repeating the steps from \ref{legendre}
\cite{tokar2019effective}. 

Similar to (\ref{V_y=U_s}) in the exact case, the equality 
\begin{equation}
v_y(y,t)=u_x(x,t) 
\label{ux=vy}
\end{equation}
holds also in LPA so at the end of renormalization this can be used to
find the equation of state in parametric form. Thus, from (\ref{y-x1})
one gets at ${t}^{\mathrm R}$ where according to (\ref{xR})
\begin{equation}
h/r = y+\bar{t}^{\mathrm R}v^{\mathrm R}_y.
\label{h-y}
\end{equation}
The expression for the magnetisation can be expressed through $y$
by replacing in (\ref{eq-of-state}) $h/r$ by the r.h.s.\ of (\ref{h-y})
which gives
\begin{equation}
m(y)=y-t_0v^{\mathrm R}_y.
\label{m-y}
\end{equation}
As is seen, $y=m$ only when $t_0=0$ in which case the equation of state
is obtained by replacing $y$ in (\ref{h-y}) by $m$.  When $t_0\not=0$
(\ref{h-y}) and (\ref{m-y}) define the equation of state parametrically.
\section{\label{irim}LPA solution of IRIM}
In the IRIM all spin pairs interact with the same dimensionless strength
$K/N$ so the sites can be numbered in an arbitrary order which means that
the model is structureless. Because the use of the conventional lattice
Fourier transform is not warranted in this case, all calculations will
be carried out in the space of the site numbers.

As is easy to see, the pair interaction matrix in IRIM can be cast
in the form
\begin{equation}
	\hat{\epsilon} = K(\hat{I}-\hat{E})
	\label{eps-infty}
\end{equation}
where the idempotent matrix $\hat{E}$, $\hat{E}^2=\hat{E}$, has matrix
elements ${E}_{ij}=N^{-1}$ and, as a consequence, matrix $\hat{I}-\hat{E}$
is also idempotent.  The spectrum of idempotent matrices consists of
only two points: 0 and 1, so the spectrum of $\hat{\epsilon}$
in (\ref{eps-infty}) is $(0, K)$ so similar to the lattice case the lowest 
eigenvalue of $\hat{\epsilon}$ is zero.

In the absence of the momentum representation the density
of states can be found according to the formula $\rho(E)=-\pi^{-1}\Im
\Tr(E+\mathrm{i}\varepsilon-\hat{\epsilon})^{-1}$.  Now using the
idempotence of $\hat{\epsilon}$ it is easy to find that
\begin{equation}
	\left(\frac{1}{z-\hat{\epsilon}}\right)_{ii}
	=\frac{1}{N}\frac{1}{z}-\left(1-\frac{1}{N}
	\right)\frac{1}{z-K}.
	\label{Gii(z)}
\end{equation}
With $z=E+\mathrm{i}\varepsilon$ one gets
\begin{equation}
	\rho(E) = N^{-1}\delta(E)+(1-N^{-1})\delta(E-K)
	\label{rho}
\end{equation}
so that according to (\ref{p})
\begin{equation}
p(t)=N^{-1}+(1-N^{-1})\theta({t^{-1}-r}-K).
\label{p-infty}
\end{equation}
As is seen, when
\begin{equation}
	0\leq t\leq t_0=1/(r+K) 
	\label{t0-infty}
\end{equation}
$p(t)=1$ and in this range the RG equation can be solved by (\ref{f_ini})
with $u(x,t_0)$ for IM (which includes IRIM as a special case) calculated
explicitly in (\ref{n=1}). By reminding that the LT
(\ref{y-x1})--(\ref{v-u1}) at $t_0$ is trivial, on the basis of
(\ref{n=1}) we obtain the explicit expression
\begin{equation}
	v(y,t_0)=\frac{y^2}{2t_0}
	-\ln\left(\cosh\frac{y}{t_0}\right)+\frac{1}{2}\ln(2\pi t_0).
	\label{v-infty}
\end{equation}
For $t>t_0$ $p(t)=1/N$ in (\ref{p-infty}) so the RG equation
(\ref{burgers}) for IRIM becomes the conventional BE with the constant
viscosity $1/2N$ and in the thermodynamic limit $N\to\infty$ it reduces to
the inviscid BE with the discontinuous shock wave solutions corresponding
to FOPTs \cite{Brankov_1983,choquard_mean_2004}. The transformed equation
(\ref{LPA2}) acquires a simple form in this limit
\begin{equation}
v_t\vert_{N\to\infty}=0.
\label{v_t=0}
\end{equation}
In this form, however, it produces the MF solution exhibiting
the undesirable van der Waals loops in the coexistence region. To see
this we note that (\ref{v_t=0}) implies that the initial condition
does not change with $t$ so at $t^{\mathrm R}=1/r$ the renormalized
$v^R(y)=v(y,t_0)$. Substituting it in (\ref{h-y}) and (\ref{m-y})
on gets after some rearrangement the MF equation of state in
parametric form
\begin{eqnarray}
h&=&y/t_0-K\tanh(y/t_0)\\
\label{h-irim}
m &=&\tanh(y/t_0)
\label{m-irim}
\end{eqnarray}
from which immediately follows the exact IRIM
MF equation \cite{salinas_introduction_2001}
\begin{equation}
	m=\tanh(Km+h).
	\label{MF}
\end{equation}
The MF free energy of IRIM \cite{salinas_introduction_2001}
\begin{equation}
	f = Km^2/2-\ln\left[2\cosh(Km+h)\right].
	\label{f-infty}
\end{equation}
is obtained from (\ref{f}) with the use of inverted transform
(\ref{v-u1}), (\ref{v-infty}), (\ref{h-irim}) and noticing that the last
term in (\ref{v-infty}) is cancelled by the last term in (\ref{f}) as is
easily calculated with the use of definitions of $\hat{G}^{\mathrm P}$
(\ref{GP}) and (\ref{eps-infty}). As is seen, the arbitrary parameter $r$
completely disappears from the solution given by equations (\ref{MF})
and (\ref{f-infty}). The MF solution, however, is not fully consistent
because FE (\ref{f-infty}) is not convex and as a consequence the van
der Waals loops will appear in the equation of state. As explained in
the main text, these deficiencies can be avoided if the thermodynamic
limit in (\ref{v_t=0}) is taken more carefully.
\section{\label{lat-vs-continuum}Particularity of the lattice LPA}
The LPA {\em ansatz} (\ref{ULPA}) in general case does not satisfy
the S-matrix EFRGE (\ref{exact0}) because of the second term on the
r.h.s.. However, the continuum and the lattice cases differ from each
other due to the difference between the Kronecker symbols in (\ref{ULPA})
which on the lattice takes the form
\begin{equation}
\delta^L_{{{\bf k}_1} +{{\bf k}_2}+\dots +{{\bf k}_l}}
=\delta_{{{\bf k}_1} +{{\bf k}_2}+\dots +{{\bf k}_l},0}
+\sum_{{\bf K}\not=0}\delta_{{{\bf k}_1} +{{\bf k}_2}+\dots 
+{{\bf k}_l,{\bf K}}}
\label{deltaL}
\end{equation}
where on the r.h.s.\ the deltas are the conventional Kronecker symbols
and in the summation over the reciprocal lattice vectors ${\bf K}$ we
singled out term ${\bf K=0}$ which is the only one that is present in
the continuum theory \cite{wilson,1984}.

Substituting (\ref{ULPA}) in the nonlinear term in (\ref{exact0}) and
using the first Kronecker delta to lift the summation over ${\bf k}$
and the definition (\ref{deltaL}) one arrives at the sum of quadratic
in $u_l$ contributions of the form
\begin{eqnarray}
&& u_l(t) u_{l^\prime}(t)\sum_{\bf k}\theta\left[ G({\bf k},t)-t\right]  
        \delta^L_{\sum_{j=1}^{l-1}\mathbf{k}_j -\mathbf{k}}
\delta^L_{\mathbf{k}+\sum_{j^\prime=1}^{l^\prime-1}
\mathbf{k^\prime}_{j^\prime}}\nonumber\\
&=&\delta^L_{\sum_{j=1}^{l-1}\mathbf{k}_j+\sum_{j^\prime=1}^{l^\prime-1}
\mathbf{k^\prime}_{j^\prime}}
u_l(t)u_{l^\prime}(t)\left.\sum_{\bf K}\theta\left[ \Lambda({\bf k}/|{\bf k}|,t)-|{\bf k}|\right]\right\vert_{{\bf k}=\sum_{j=1}^{l-1}\mathbf{k}_j -\mathbf{K}}
        \label{quadratic_term}
\end{eqnarray}
where $\Lambda$ in the argument of the step function on the second line
should be determined from equation (\ref{omega}) defining the momentum
cutoff. In the lattice case it depends on the crystal anisotropy, i.e., on
the momentum direction ${\bf k}|{\bf k}|$. In the isotropic case $\Lambda$
would coincide with the conventional momentum cutoff \cite{wilson,1984}.

If the step functions in (\ref{quadratic_term}) were equal to unity
(the LPA) than the terms (\ref{quadratic_term}) would sum up to the
second term on the l.h.s.\ in (\ref{LPA}). But this would be valid
only if the arguments of the step functions were always positive. In
the isotropic continuum theory when only ${\bf K=0}$ contributes to
(\ref{quadratic_term}) this would mean that the sum of ${\bf k}_j$ is
smaller in absolute value than $\Lambda$. But this can be guaranteed
only for $l\leq2$ because, by our definition of the renormalization
procedure, all individual field momenta reside within $\Omega(t)$ defined
in (\ref{omega}).  However, for $l>2$ the absolute value of the sum may
exceed $\Lambda$ in which case the step functions will dependent on ${\bf
k}_j$. As a result, many contribution in the momenta space will be lost
from contributions $l,l^\prime>2$ and the errors introduced by the LPA
will be enhanced in the strong coupling case when $u_{l>2}$ are large.

The problem alleviates in the lattice models when all ${\bf K}$ contribute
in (\ref{quadratic_term}). This can be visualized by considering the
step functions in the periodic zone scheme where each ${\bf K}$ becomes
the centre of a region $\Omega_{\bf K}$ which is the same $\Omega$ as in
(\ref{omega}) only shifted on vector ${\bf K}$. Now the step function
in (\ref{quadratic_term}) is equal to unity within all $\Omega_{\bf
K}$ and zero at the outside. In the course of the renormalization
this will create a muffin-tin structure where at the early stages of
RG flow there will much more volume in momentum space where the LPA
conjecture is fulfilled. For example, in the region $0\leq t\leq t_0$
in figure \ref{figure1} $\Omega=$~BZ so in the periodic zone scheme
the whole momentum space becomes covered and the step function is unity
everywhere.  This makes LPA exact which has allowed us to perform the
exact renormalization in (\ref{f_ini}) and (\ref{f_ini-n}). As $t$
grows above $t_0$ all $\Omega_{\bf K}$ start to shrink but close to
$t_0$ only narrow gaps between $\Omega_{\bf K}$ will appear so the
violation of the LPA assumption $\theta=1$ in (\ref{quadratic_term})
will be relatively small and can be roughly assessed by the value of
$\kappa(t)={\rm vol}[\bar{\Omega}(t)]/{\rm vol}[{\Omega}(t)]$.  We note
that the estimate is not based on the smallness of the momentum, as in
DE, or on the values of $u_l$. In fact, it relies on the closeness of
$\Omega\simeq$ to BZ while $|{\bf k}|$ may be as large as $\sim\sqrt{3}\pi$
on sc lattice.  As $t$ advances farther toward $t^{\mathrm R}$ $\Omega$
shrinks and $\kappa(t)$ grows to large values which signals the
breakdown of the LPA assumption (\ref{G_t=1}) in a large part of BZ.
However, because under Wilsonian renormalization $u_l$ with large $l$
fast attenuate towards small values \cite{wilson}, it is expected that
the terms quadratic in $u_l$ violating the LPA assumption will become
small. Besides, closer to $t^{\mathrm R}$ $\Omega(t)$ tends to be almost
spherical, the lattice nature of the system smears out and $k/\Lambda$
becomes an acceptable DE parameter to justify LPA on the late stages of
renormalization. It is to be stressed that the above reasoning relied
on the shape of the cutoff function similar to the step function and
is not applicable to the Wilson-type momenta elimination in large-$k$
shells \cite{wilson}. This seems to be confirmed by the RG calculations in
\cite{CAILLOL2012291} where the use of the Wilson-type cutoff led to poor
agreement with the MC simulations at large interaction strengths while the
use of the step function resulted in a perfect agreement with the MC data.
\section{\label{ini}Initial condition in the spin models}
In $O(n)$-symmetric case the $n$-dimensional integral in the exact initial
renormalization (\ref{f_ini}) takes the form
\begin{equation}
	\rme^{-u({\bf x},t_0)}
=\frac{1}{(2\pi t_0)^{n/2}}\int d{\bf x}_0\,
\exp{\left(-\frac{({\bf x-x}_0)^2}{2t_0}\right)}\rme^{-u({\bf x}_0,0)}.
\label{f_ini-n}
\end{equation}
For spin-lattice models we assume that for all $n$ the spin length is
equal to unity
\begin{equation}
\rme^{-u({\bf x}_0,0)}=\delta({\bf x}_0^2-1).
\label{s=1}
\end{equation}
The integration in (\ref{f_ini-n}) is conveniently carried out in
hyperspherical coordinates \cite{n-sphere} in which the integration
over $|{\bf x}_0|$ is trivial due to the delta-function in (\ref{s=1})
and with the choice of the direction of ${\bf x}$ along the first axis: 
${\bf x}=(x\cos\vartheta,0,0,\dots,0)$, $x=|{\bf x}|$ (\ref{f_ini-n})
reduces to
\begin{eqnarray}
	&&\rme^{-u^{(n)}({\bf x},t_0)}
=C_n\rme^{-\frac{t_0a^2}{2}}\int_0^{\pi}
\rme^{\frac{x}{t_0}\cos\vartheta}\sin^{n-2}\vartheta\,d\vartheta\nonumber\\
&&=C_n\rme^{-\frac{t_0a^2}{2}}\int_{-1}^1\rme^{az}
(1-z^2)^{\frac{n-3}{2}}\;dz
\equiv C_n\rme^{-\frac{t_0a^2}{2}}b(n,a)
\label{theta}
\end{eqnarray}
where in $C_n$ are gathered all ${\bf x}$-independent factors from
(\ref{f_ini-n}) and from the spherical volume \cite{n-sphere} which can
be easily recovered if needed; $a=x/t_0$, the integral on the second
line is obtained by the change of variables $z=\cos\vartheta$. Functions
$b(n,a)$ have been introduced in order to make use of a recursion relation
they satisfy. The latter is obtained by integrating by parts twice in
(\ref{theta}) with respect to $d(\rme^{az})$ to get
\begin{equation}
b\left(n,a\right)=\frac{n-3}{a^2}
\left[(n-5)b\left(n-4,a\right)
-(n-4)b\left(n-2,a\right)\right].
\label{recursion}
\end{equation}
As is seen, functions $b(n,a)$ can be calculated recursively for all
$n\geq6$ if they are known for $n=2-5$.  The latter are given by the terms
in brackets in the expressions calculated directly from (\ref{theta}):
\begin{eqnarray}
\rme^{-u^{(2)}({\bf x},t_0)}
&=&C_2\rme^{-\frac{t_0a^2}{2}}\left[\pi I_0\left(a\right)\right]
\nonumber\\
\rme^{-u^{(3)}({\bf x},t_0)}
&=&C_3\rme^{-\frac{t_0a^2}{2}}\left[\frac{2}{a}
\sinh a\right]
\nonumber\\
\rme^{-u^{(4)}({\bf x},t_0)}
&=&C_4\rme^{-\frac{t_0a^2}{2}}\left[
\frac{\pi}{a}I_1\left(a\right)\right]
\nonumber\\
\rme^{-u^{(5)}({\bf x},t_0)}
&=&C_5\rme^{-\frac{t_0a^2}{2}}\left[\frac{4}{a^3}
\left(a\cosh a-\sinh a\right)\right]
	\label{n=2-5}
\end{eqnarray}
where $I_{0,1}$ are the modified Bessel functions of the first kind.

Case $n=1$ is not covered by (\ref{theta}) but using (\ref{s=1})
multiplied by 2 to accord with the conventional definition of IM, the
partially renormalized potential can be straightforwardly calculated as
\begin{equation}
	u^{(1)}({x},t_0)
	=\frac{t_0a^2}{2}-\ln(2\cosh a) + \frac{1}{2}\ln (2\pi t_0)
\label{n=1}
\end{equation}
where we took into account field-independent terms needed in \ref{irim}.
\providecommand{\newblock}{}

\end{document}